\documentclass[fleqn,usenatbib]{mnras}

\usepackage{newtxtext,newtxmath}

\usepackage[T1]{fontenc}

\DeclareRobustCommand{\VAN}[3]{#2}
\let\VANthebibliography\thebibliography
\def\thebibliography{\DeclareRobustCommand{\VAN}[3]{##3}\VANthebibliography}

\usepackage{graphicx}	
\usepackage{amsmath}	

 \newcommand\kms
 {km s$^{-1}$}

\title[AGATE: Virgo Kinematics]{Analysis of Galaxies at the Extremes: A Kinematic Analysis of the Virgo Cluster Dwarfs VCC~9 and VCC~1448 using the Keck Cosmic Web Imager}

\author[J. S. Gannon et al.]{
Jonah S. Gannon,$^{1,2}$\thanks{E-mail: jonah.gannon@gmail.com}
Duncan A Forbes,$^{1,2}$ 
Aaron J. Romanowsky,$^{3,4}$
Jean P. Brodie,$^{1,2,4}$
Lydia Haacke,$^{1,2}$
\newauthor{Anna Ferr\'e-Mateu,$^{5,6,1}$
Shany Danieli$^{7}$, 
Pieter van Dokkum,$^{8}$
Maria Luisa Buzzo,$^{1,2}$
Warrick J. Couch,$^{1,2}$}
\newauthor{and Zili Shen$^{8}$}
\\
$^{1}$ Centre for Astrophysics and Supercomputing, Swinburne University, John Street, Hawthorn VIC 3122, Australia \\
$^{2}$ ARC Centre of Excellence for All Sky Astrophysics in 3 Dimensions (ASTRO 3D) \\
$^{3}$ Department of Physics and Astronomy, San Jos\'e State University, One Washington Square, San Jose, CA 95192, USA \\
$^{4}$ Department of Astronomy \& Astrophysics, University of California Santa Cruz, CA 95064, USA \\
$^{5}$ Instituto de Astrof\'isica de Canarias, Calle V\'ia L\'actea S/N, E-38205, La Laguna, Tenerife, Spain\\
$^{6}$ Departamento de Astrofísica, Universidad de La Laguna, 38206, La Laguna (S.C. Tenerife), Spain\\
$^{7}$ Department of Astrophysical Sciences, 4 Ivy Lane, Princeton University, Princeton, NJ 08544\\
$^{8}$ Department of Astronomy, Yale University, New Haven, CT 06511, USA}

\date{Accepted XXX. Received YYY; in original form ZZZ}

\pubyear{2023}

\begin{document}
\label{firstpage}
\pagerange{\pageref{firstpage}--\pageref{lastpage}}
\maketitle

\begin{abstract}
We present spatially resolved Keck Cosmic Web Imager stellar spectroscopy of the Virgo cluster dwarf galaxies VCC~9 and VCC~1448. These galaxies have similar stellar masses and large half-light radii but very different globular cluster (GC) system richness ($\sim$25 \textit{vs.} $\sim$99 GCs). Using the KCWI data, we spectroscopically confirm 10 GCs associated with VCC 1448 and one GC associated with VCC 9. We make two measurements of dynamical mass for VCC~1448 based on the stellar and GC velocities respectively. VCC 1448’s mass measurements suggest that it resides in a halo in better agreement with the expectation of the stellar mass – halo mass relationship than the expectation from its large GC counts.. For VCC~9, the dynamical mass we measure agrees with the expected halo mass from both relationships. We compare VCC~1448 and VCC~9 to the GC-rich galaxy Dragonfly~44 ($\sim74$~GCs), which is similar in size but has $\sim 1$~dex less stellar mass than either Virgo galaxy. In dynamical mass -- GC number space, Dragonfly~44 and VCC~1448 exhibit richer GC systems given their dynamical mass than that of VCC~9 and other `normal' galaxies. We also place the galaxies in kinematics--ellipticity space finding evidence of an anticorrelation between rotational support and the fraction of a galaxy's stellar mass in its GC system. i.e., VCC~9 is more rotationally supported than VCC~1448, which is more rotationally supported than Dragonfly~44. This trend may be expected if a galaxy's GC content depends on its natal gas properties at formation.
\end{abstract}




\section{Introduction}

Large, low surface brightness (LSB) galaxies in the dwarf-like regime have been studied for their peculiarities since at least the 1950's (see e.g., \citealp{Reaves1953, Reaves1956}). For example, large-sized, low-surface brightness cluster galaxies were studied by \cite{2003AJ....125...66C} and it was argued they must be dark matter dominated in order to survive in the cluster environment \citep{2009MNRAS.393.1054P}. However, it is only recently that widespread interest in the study of these extreme galaxies has been renewed with a sub-set of LSB galaxies, so-called ``ultra-diffuse" galaxies \citep[UDGs]{vanDokkum2015}. It soon became apparent that many are also extreme in their globular cluster (GC) content \citep{vanDokkum2017}. These galaxies may also help us differentiate between different prescriptions of dark matter (see e.g., \citealp{Wasserman2019}), with understanding star formation at the extreme (see e.g., \citealp{Kadofong2022a, Kadofong2022b}) as well as understanding the first epochs of galaxy and GC formation \citep{Peng2016, Forbes2020, Danieli2022}. 

There have been various attempts to try and explain the formation of such large-sized dwarf galaxies. These attempts have focused on formation via internal (e.g., star formation feedback \citealp{DiCintio2017, Chan2018}; or high-halo spin \citealp{Amorisco2016, Rong2017, Benavides2023}) or external (e.g., tidal heating \citealp{Carleton2019}; tidal stripping \citealp{Sales2020, Doppel2021}; or galaxy collisions/mergers \citealp{Wright2021, vanDokkum2022, Gannon2023b}) physical processes. Combinations of both are also possible (e.g., \citealp{Jiang2019, Martin2019, Sales2020}). While it is clear that large dwarf galaxies likely have multiple pathways to their formation (e.g., \citealp{Papastergis2017}; \citealp{Buzzo2022}; \citealp{FerreMateu2023}, \citealp{Buzzo2024}), understanding the individual pathways is still of interest. In particular, it has been shown that measuring the dynamics of these galaxies can be key to differentiate which formation pathways are viable (\citealp{Beasley2016, Toloba2018, Toloba2023, vanDokkum2019b, Gannon2020, Gannon2021, Gannon2022, Gannon2023a}).

Furthermore, the large variety of GC numbers the galaxies host, are suggestive of dual formation pathways \citep{Forbes2020}. Specifically, assuming that these galaxies follow the GC-number -- halo mass relation \citep{Burkert2020}, the wide range of observed GC-richness implies a large spread in the total dark matter halo masses in which they reside. In turn, this would imply a wide range of formation pathways. Crucially, this is only true if the galaxies follow the same GC number -- halo mass relationship as normal galaxies. Building a sample of large, GC-rich dwarf galaxies with independent halo mass measurements (e.g., \citealp{vanDokkum2019b}) is vital to test this assumption \citep{Forbes2024}. 

Beyond independent halo mass measurements, resolved kinematics of normal galaxies have aided our understanding of their formation (c.f., the review of \citealp{Cappellari2016}). For example, the build-up of stellar mass in a galaxy is thought to leave imprints in their resolved kinematics. Galaxies built via the accretion of high angular momentum gas are expected to rotate faster than galaxies built through randomised mergers, which tend to be more dispersion-supported in their kinematics. However, some orientations of major mergers are known to lead to higher angular momentum/rotation (e.g., \citealp{Wright2021}). These differences have led multiple authors to classify galaxies as either ``slow rotators" or ``fast rotators" based on metrics relying on the ratios of rotation to dispersion support in their {\it central} kinematics (e.g., \citealp{Emsellem2011, Arnold2014, FraserMckelvie2022}). Historically, kinematically resolved studies of galaxies have mainly relied on the use of long-slit spectroscopy, but recent developments in integral field spectroscopy have allowed far more complete galaxy samples to be compiled. See for example, the ATLAS-3D project \citep{Cappellari2011a}, the CALIFA project \citep{Sanchez2012}, the SAMI project \citep{Bryant2015}, the MANGA project \citep{Bundy2015} or the LEWIS project \citep{Iodice2023}. 


Here, we target two large Virgo cluster dwarf galaxies --  VCC~1448 (IC~3475; $R_{\rm e} \approx 3.5$~kpc and $M_{\star}\approx2.6\times10^9$~M$_{\odot}$) and VCC~9 (IC~3019; $R_{\rm e} \approx 3$~kpc and $M_{\star}\approx2.5\times10^9$~M$_{\odot}$) using resolved kinematics from the Keck Cosmic Web Imager (KCWI; \citealp{Morrissey}), an integral field unit on the Keck II telescope. VCC~1448 is prototypical for a large-LSB dwarf, with \citet{Reaves1956} first cataloguing it in a collection of large, LSB objects in the Virgo cluster as ``IC~3475 type objects". VCC~9 is of comparable size and stellar mass. A key point of difference between the two galaxies is their GC content, and hence expected total halo mass. VCC~1448 is particularly rich in GCs with ground-based imaging suggesting a total system of 99.3$\pm$17.6 GCs \citep{Lim2020}, while VCC~9 is more representative of a galaxy with its luminosity, hosting 25.7$\pm$6.4~GCs \citep{Peng2008}. Likewise, we will also contrast these galaxies with the Coma Cluster UDG Dragonfly 44. Dragonfly 44 is of similar size and GC richness ($N_{\rm GC}=74\pm18$, \citealp{vanDokkum2017}) to VCC~1448, however has an order of magnitude less stellar mass ($M_{\star}\approx3\times10^8$~M$_{\odot}$). Dragonfly~44 has been proposed to be prototypical for a ``failed galaxy'' - a galaxy that quenched catastrophically before forming the total stellar mass expected for its dark matter halo \citep{vanDokkum2015, Peng2016, Webb2022}. Currently, Dragonfly~44 is the only ``failed galaxy'' UDG with resolved kinematics measured. However, many other examples of UDGs with similar GC systems exist (e.g., NGC~5846-UDG1; \citealp{Forbes2019, Danieli2022}) making elucidating their formation vital. 

This paper aims to focus on the comparative kinematic differences between VCC~9 and VCC~1448. Furthermore, given the known trends of luminosity/half-light radius and GC-richness (see e.g., \citealp{Harris2013}), we also aim to understand the underlying galaxy structure that may induce a different GC formation efficiency at fixed size/luminosity. We undertake this work as part of the Analysis of Galaxies at the Extremes (AGATE) project. In this project, we seek to better understand extreme galaxies with a particular emphasis on those galaxies that are at the size or GC-richness extremes for their stellar mass.

In Section \ref{sec:targets} we provide further details of the two targets. In Section \ref{sec:kcwi_data} we outline our KCWI data, describing their observation, reduction and presenting the results of our analysis. We discuss these results in the context of other Virgo cluster dwarf galaxies and contrast VCC~1448 with Dragonfly 44 in Section \ref{sec:discussion}. Finally, we present our primary conclusions in Section \ref{sec:conclusions}.

\section{Targets} \label{sec:targets}
\begin{figure}
    \centering
    \includegraphics[width = 0.48 \textwidth]{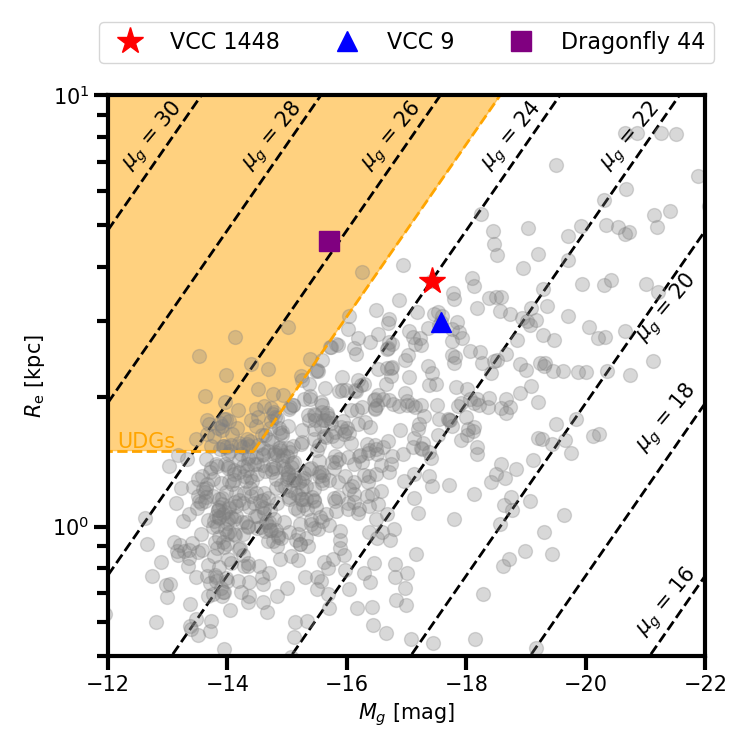}
    \caption{Semi-major half-light radius \textit{vs.} $g$-band absolute magnitude. We plot grey points for Virgo cluster dwarf galaxies from the catalogue described in \citet{Weinmann2011} and \citet{Lisker2013}. The positions within this parameter space of VCC~1448 and VCC~9 are indicated by a red star and blue triangle respectively.  We include the region matching the UDG definition of $R_{\rm e} > 1.5$ kpc and $\langle\mu_{g}\rangle_{\rm e} > 25$ mag arcsec$^{-2}$ in orange. In it, we place the UDG Dragonfly 44 (purple square; \citealp{vanDokkum2015}) which will become a relevant galaxy of comparison in Section \ref{sec:discussion}. Under the assumption of circular galaxies, lines of constant surface brightness are indicated with dashed lines. Units for these are mag arcsec$^{-2}$. Both VCC~1448, VCC~9 and Dragonfly 44 are amongst the largest galaxies for their luminosity in the Virgo sample. }
    \label{fig:re_lum_context}
\end{figure}

\begin{figure}
    \centering
    \includegraphics[width = 0.48 \textwidth]{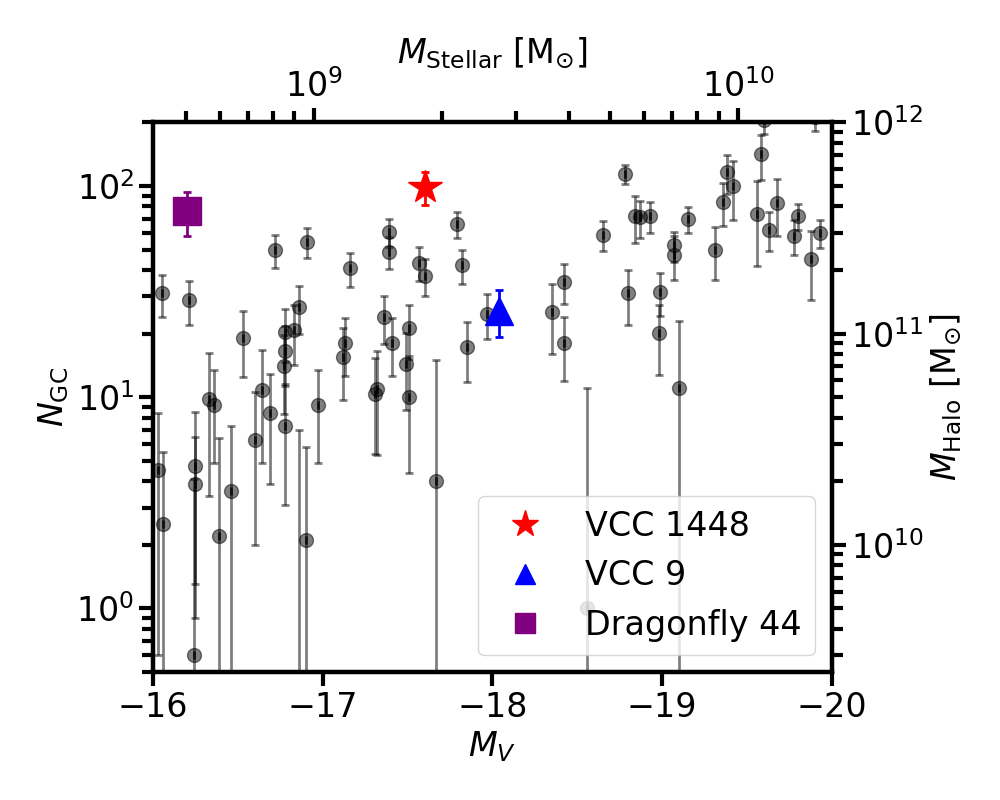}
    \caption{GC-number \textit{vs.} $V$-band absolute magnitude. We additionally include a second x-axis converting $V$-band magnitude to stellar mass assuming $M_\star/L_V=2$ and a second $y$-axis converting GC-numbers into $M_{\rm Halo}$ using the \citet{Burkert2020} relationship. We plot Virgo cluster galaxies with data from the \textit{HST} ACSVCS (grey points; \citealp{Peng2008}). We highlight VCC~9 (blue triangle) from \citet{Peng2008}, VCC~1448 (red star) using data from \citet{Lim2020} and Dragonfly 44 (purple square) using the GC counts of \citet{vanDokkum2017}. While VCC~9 has a normal GC content for its luminosity, VCC~1448 and Dragonfly 44 exhibit rich GC systems more characteristic of galaxies $\sim 2$ magnitudes brighter. This may be indicative of an unusually massive dark matter halo for their stellar mass, as indicated by the second axes.}
    \label{fig:gc_lum_context}
\end{figure}

In this work, we study two Virgo cluster galaxies, VCC~9 and VCC~1448, and make comparisons to the Coma Cluster UDG Dragonfly 44. It is worth knowing that while VCC~1448 has already had some of its GC kinematics studied by \citet{Toloba2023} our work is the first time its stellar kinematics have been studied. We contextualise our choice to study these galaxies in Figures \ref{fig:re_lum_context} and \ref{fig:gc_lum_context}. 

In Figure \ref{fig:re_lum_context} we plot the catalogue described in \citet{Weinmann2011} and \citet{Lisker2013} in half-light radius -- absolute magnitude space, with both galaxies highlighted. The region commonly assigned to the so-called UDGs \citep{vanDokkum2015} is highlighted in orange. VCC~9 and VCC~1448 are amongst the largest dwarf galaxies in the catalogue for their luminosity. Indeed, VCC~1448 has been previously identified by numerous authors as being an outlier to large sizes for its luminosity \citep{Reaves1956, Binggeli1985, Lim2020}. While neither fit the UDG definition, it is likely both fit definitions based on galaxies being outliers in size -- stellar mass space such as the ``ultra-puffy galaxy'' definition proposed by \citet{Li2023}. Dragonfly 44 is also extremely large for its luminosity and resides in the region corresponding to UDGs.

In Figure \ref{fig:gc_lum_context} we plot GC counts \textit{vs.} absolute magnitude. We use these two parameters as observational proxies for the stellar mass -- halo mass relationship by converting luminosity into stellar mass assuming a stellar mass to light ratio ($M_\star/L_V$) of 2 (see e.g., \citealp{vanDokkum2018}) and converting GC-numbers into a halo mass ($M_{\rm Halo}$) using the \citet{Burkert2020} relationship. The GC counts for VCC~9 (25.7 $\pm$ 6.4; \citealp{Peng2008}) are typical for a galaxy at its luminosity. However, the GC estimate for VCC~1448 (99.3$\pm$17.6; \citealp{Lim2020}) is elevated for its luminosity. Based on an average Milky Way GC mass of $\sim2\times10^{5}$~$\mathrm{M_{\odot}}$~(\citealp{Harris1996}; 2010 revision), the GC system represents $\sim 0.76\%$~of its total stellar mass. For VCC~9 the GC system is only 0.21\%~ of its stellar mass.


As can be seen in Figure \ref{fig:re_lum_context}, Dragonfly~44 has a similar size to VCC~1448 but is 1.7 magnitudes fainter ($-15.7$ vs. $-17.4$ mag.). Estimates of its GC-richness have Dragonfly~44 as extremely GC-rich for its luminosity (Figure \ref{fig:gc_lum_context}; although see \citealp{Saifollahi2022} and \citealp{Forbes2024} for a discussion of Dragonfly~44's GC richness). Due to this high number of GCs and comparatively low luminosity, Dragonfly 44 and other GC-rich UDGs have been thought of as ``failed galaxies" or ``pure stellar haloes" \citep{Peng2016}. Given the similarly large size and rich GC system, the primary difference between VCC~1448 and Dragonfly 44 is the $\sim1$~dex difference in stellar mass. We therefore wish to test if both reside in similar dark matter halos to probe for an evolutionary link. i.e., we wish to test if Dragonfly 44 could plausibly be a passively evolved version of VCC~1448 that began on a similar evolutionary path or, equivalently, if VCC~1448 may be what Dragonfly~44 could have evolved into without `failing'.

In Table \ref{tab:target_Summary} we provide a basic summary of the relevant parameters for each galaxy. For VCC~1448 the values for rows 1--7 are taken from \citet{Ferrarese2020} and \citet{Toloba2023}. The GC counts in row 8 are from \citet{Lim2020}. Rows 9--16 are from this work. For VCC~9 the values for rows 1--4 and 6 are taken from the catalogue of \citet{Weinmann2011} and \citet{Lisker2013}. Row 7, the GC counts, are from \citet{Peng2008}. All other values are from either \citet{Toloba2014} and \citet{Toloba2015} or are derived here based on their work. In particular, we do not use their stated dynamical mass for VCC~9 but instead re-derive their dynamical mass in Section \ref{sec:dynamical_masses}. For Dragonfly 44 rows 1--7 are taken from \citet{vanDokkum2015} and \citet{vanDokkum2017}. Row 8 is calculated in this work. The remaining rows are from \citet{vanDokkum2019b}.

\begin{table}
    \centering
    \begin{tabular}{llll}
    \hline
    Quantity & VCC~1448 & VCC~9 & Dragonfly~44  \\ \hline
    RA [J2000] & 188.17001 & 182.34271 & 195.24167 \\
    Dec. [J2000] & +12.77108 & +13.99243 & +26.97639  \\
    $D$ {[}Mpc{]} & 16.5 & 16.5 & 100 \\
    $M_g$ {[}mag{]} & $-$17.4 & $-$17.6 & $-$15.7 \\
    $M_{\star}$ [M$_{\odot}$] & $2.6 \times 10^{9}$ & $2.5 \times 10^{9}$ & $3 \times 10^{8}$\\
    $R_{\rm e}$ {[}''{]} / {[}kpc{]} & 43.92 / 3.51 & 37.13 / (2.97) &  9.69 / (4.7) \\
    $\epsilon$ & 0.19 & 0.19 & 0.35 \\
    $N_{\rm GC}$ & 99.3 (17.6) & 25.7 (6.4) & 74 (18) \\
    $M_{\rm GC}/M_{\star}$ & 0.76\% & 0.21\% & 5.07\% \\
    $V_{R}$ {[}km s$^{-1}${]} & 2280 (6) & 1674 (6) & 6234 (-) \\
    $\langle \sigma_{\rm e} \rangle$ {[}km/s{]} & 24.0 (1.6) & 26.0 (4.6) & 33 (3) \\
    $\langle V_{\rm rot, e} \rangle$ [km/s] & 9.4 (1.0) & 20.2 ($^{+4.8} _{-5.2}$) & 0$^\dagger$ \\
    $\langle V \rangle /\langle \sigma \rangle$ & 0.39 (0.07) & 0.77 & 0 \\
    $M_{\rm Dyn}$ (stars) [$\times10^{9}$ M$_\odot$] & 1.96 (0.68) & 2.69 (2.29) & 3.9 (0.5)\\
    \hline
    \end{tabular}
    \caption{The basic properties of the galaxies that are compared in this work. When available uncertainties are given in parentheses. From top to bottom the rows are: 1) right ascension, 2) declination, 3) assumed distance, 4) $g$-band absolute magnitude, 5) stellar mass, 6) half-light radius, 7) ellipticity, 8) GC number, 9) percentage of stellar mass in the GC-system, 10) recessional velocity, 11) Flux-weighted average stellar velocity dispersion within the half-light radius, 12) Flux-weighted average rotation velocity within the half-light radius, 13) the average rotation/velocity dispersion within the half-light radius, and 14) Dynamical mass calculated from the stellar motions. $\dagger$ This value is assumed to be zero based on the lack of rotation seen in table 2 of \citet{vanDokkum2019b}. It is also consistent with the maximum rotation value quoted in section 5.1 of \citet{vanDokkum2019b}. For VCC~1448, rows 9--14 are from this work. For other values see the literature references at the end of Section \ref{sec:targets}. }
    \label{tab:target_Summary}
\end{table}

\begin{figure}
    \centering
    \includegraphics[width = 0.48 \textwidth]{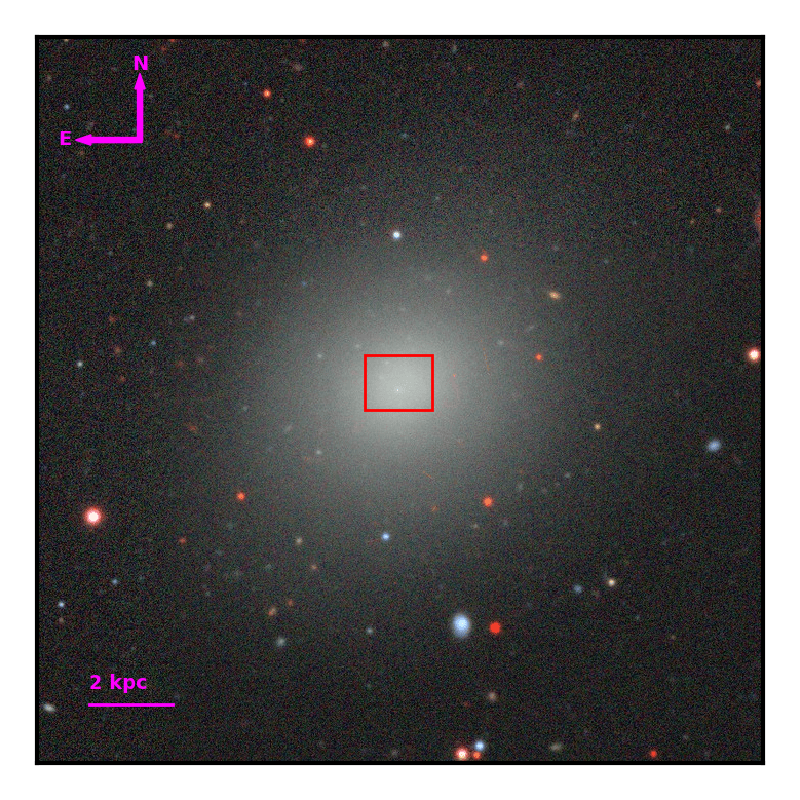}
    \caption{The on-sky positioning of our KCWI data for VCC~9. We show a 3.5' $\times$ 3.5' colour cutout centred on VCC~9 taken from the DECaLS Skyviewer (\url{https://www.legacysurvey.org/viewer}). The position of our KCWI Medium/BM/5075 slicer pointing targeting VCC~9 is indicated with a red rectangle. A line indicating 2 kpc at the Virgo cluster distance is on the bottom left. North is up and East is left as indicated.}
    \label{fig:vcc9_on_sky}
\end{figure}

\begin{figure}
    \centering
    \includegraphics[width = 0.48 \textwidth]{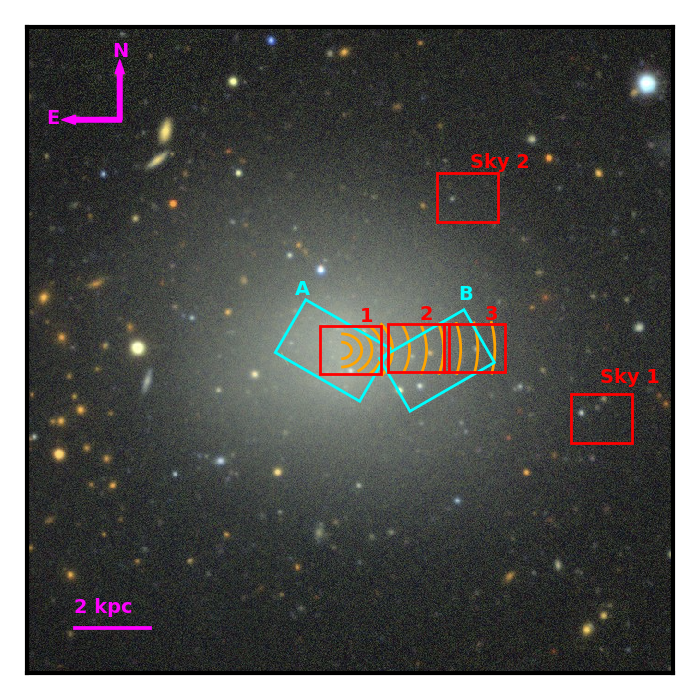}
    \caption{The on-sky positioning of our KCWI data for VCC~1448. We show a 3.5' $\times$ 3.5' colour cutout centred on VCC~1448 taken from the DECaLS Skyviewer (\url{https://www.legacysurvey.org/viewer}). The positions of our 3 KCWI Medium/BM/5075 slicer pointings targeting VCC~1448 and two offset skies are indicated with red rectangles. The positions of our two KCWI Large/BH3/5080 pointings are indicated with cyan rectangles. We display the borders of apertures used for the extraction described in Section \ref{sec:vcc1448_profile} as orange arcs. A line indicating 2 kpc at the Virgo cluster distance is on the bottom left. North is up and East is left as indicated.}
    \label{fig:on_sky}
\end{figure}

\section{Keck Cosmic Web Imager Data} \label{sec:kcwi_data}

\begin{figure*}
    \centering
    \includegraphics[width = 0.98 \textwidth]{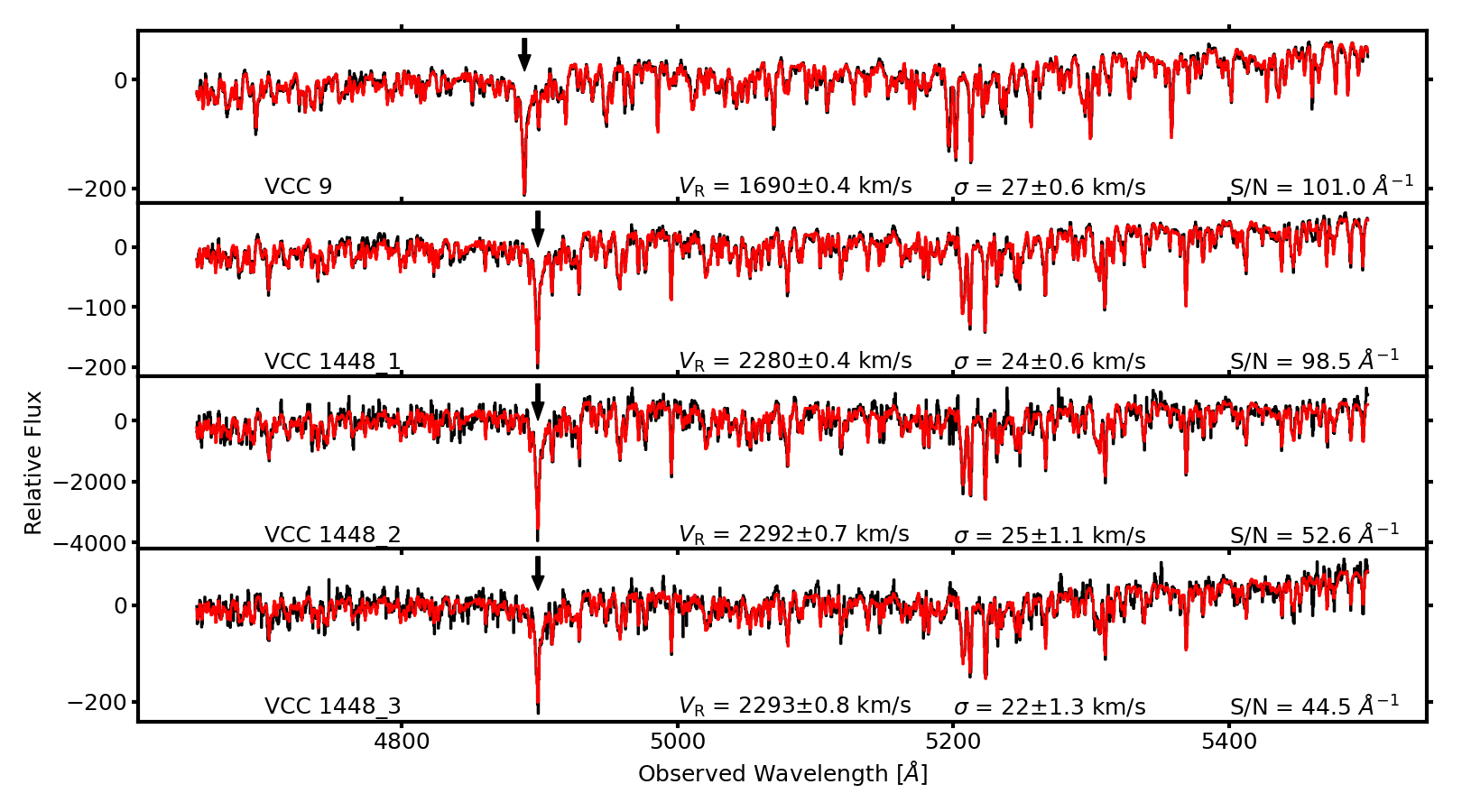}
    \caption{Examples of the Keck spectra derived from our data. We plot the final spectra from each of the four KCWI medium pointings targeting VCC~9/VCC~1448 after masking bright objects and collapsing over the entire spatial region (all in black). From top to bottom these spectra are for: the central KCWI slicer positioned on VCC~9, the first (inner) slicer positioned on VCC~1448, the second (middle) slicer positioned on VCC~1448 and the third (outer) slicer positioned on VCC~1448. These are as labelled in Figure \ref{fig:on_sky}. A best fitting \texttt{pPXF} fit is shown over each in red. The positioning of H$\beta$ is indicated for each spectrum with a black arrow. Based on fits such as these we measure recessional velocities and velocity dispersions as are labelled in each subplot.}
    \label{fig:spectra}
\end{figure*}

\begin{figure}
    \centering
    \includegraphics[width = 0.48 \textwidth]{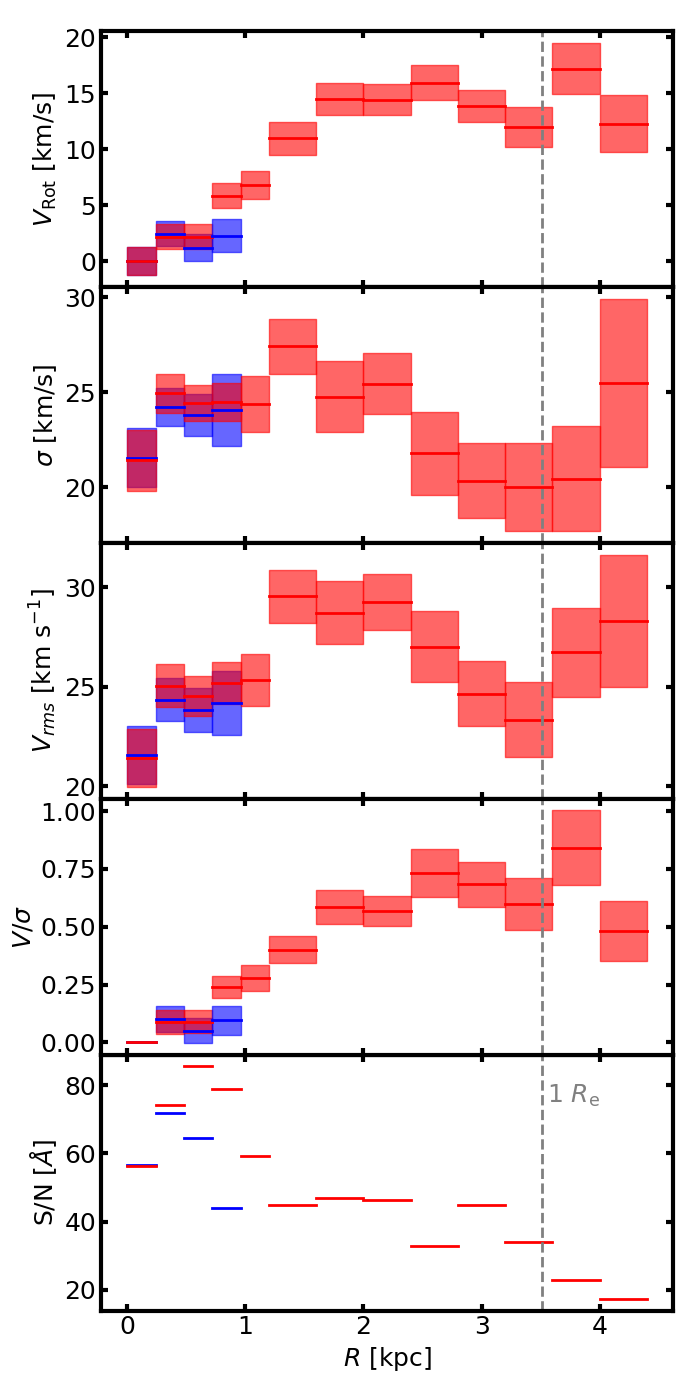}
    \caption{The derived radial stellar kinematic profiles for VCC~1448 along the semi-major axis as described in Section \ref{sec:vcc1448_profile}. Measurements along the major axis to the west are red solid lines with 1-$\sigma$ uncertainties as the red shading. Measurements out to the east along the major axis are in blue. For the rotational velocity, the values out to the east have had their negative sign dropped when plotting. From top to bottom the panels are as follows. \textit{First Panel:} The rotation profile with azimuthal correction applied. \textit{Second Panel:} The velocity dispersion profile. \textit{Third Panel:} The combined rotation plus dispersion, $V_{\rm rms}$, profile. \textit{Fourth Panel:} The $V/\sigma$~ profile. \textit{Fifth Panel:} The signal-to-noise ratio of the spectra used in the profile. In each panel we mark the half-light radius of VCC~1448 with a dashed vertical grey line.}
    \label{fig:radial}
\end{figure}

\begin{figure}
    \centering
    \includegraphics[width = 0.47 \textwidth]{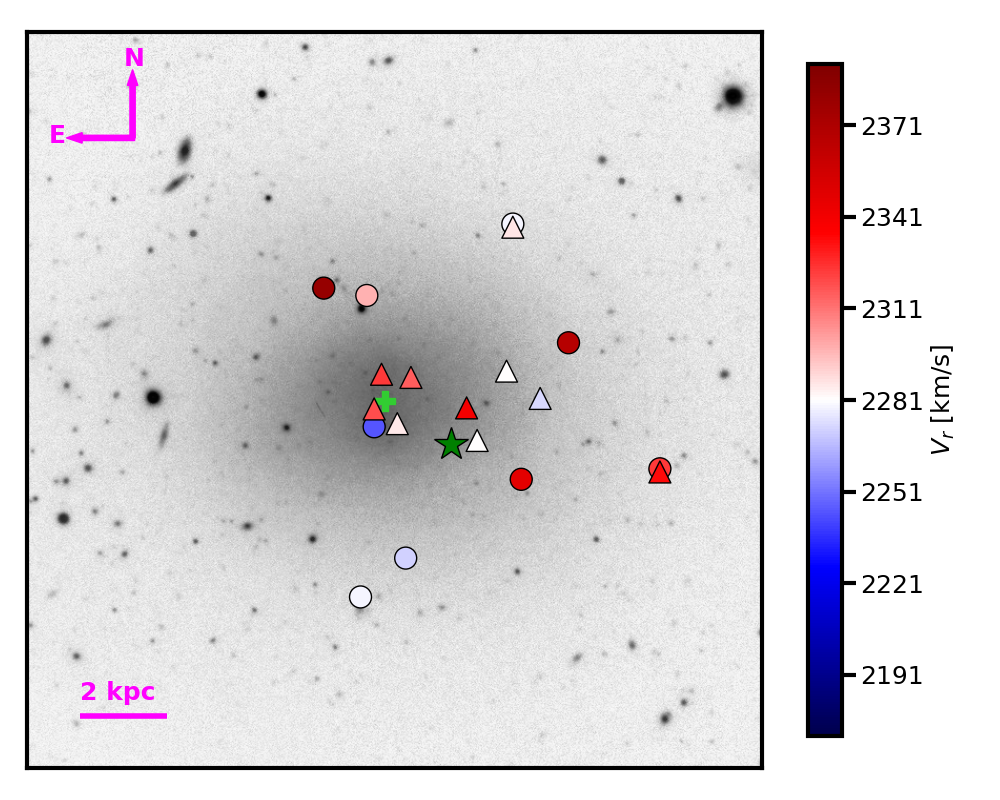}
    \caption{The on-sky recessional velocities of the confirmed VCC~1448 GCs. Compact sources we have confirmed as GCs have coloured triangles with colouring indicating their recessional velocity with respect to VCC~1448. Coloured circles indicate GCs confirmed by \citet{Toloba2023} in the same manner. Two GCs are included in both studies. An object believed to be a Milky Way star based on its recessional velocity is indicated with a green star. The centre of the galaxy is marked as a green plus. A line indicating 2 kpc at the Virgo cluster distance is on the bottom left. North is up and East is left as indicated. The majority of the GCs are colour-coded red as their recessional velocities are above that of VCC~1448 (2281 \kms). This is partially due to the spatial sampling of our KCWI data being along an axis that is rotating away from us (see Figure \ref{fig:radial}). }
    \label{fig:gcs_onsky}
\end{figure}

\begin{table*}
    \centering
    \begin{tabular}{ccccccccc}
    \hline
    Identifier & RA & Dec. & $m_{\rm F555W}$ & $m_{\rm F814W}$ & $D_{\rm VCC 1448}$ & $D_{\rm VCC 1448}$ & $V_{\rm R}$ & S/N \\
     & [J2000] & [J2000] & [mag] & [mag] & [arcsec] & [kpc] & [km s$^{-1}$] & [\AA$^{-1}$] \\ \hline
    GC 1$^*$ & 12:32:35.45 & +12:45:55.26 & 21.772 & 21.215 & 81.0 & 6.48 & 2334 (6) & 24 \\
    GC 2 & 12:32:37.80 & +12:46:16.69 & 22.094 & 21.575 & 43.92 & 3.5 & 2273 (7) & 10 \\
    GC 3$^*$ & 12:32:38.33 & +12:47:05.65 & 22.666 & 22.014 & 61.5 & 4.92 & 2287 (8) & 6 \\
    GC 4 & 12:32:38.45 & +12:46:25.11 & 22.937 & 22.496 & 35.6 & 2.85 & 2282 (8) & 12 \\
    GC 5 & 12:32:39.23 & +12:46:14.34 & 22.496 & 21.992 & 23.1 & 1.84 & 2341 (7) & 9 \\
    GC 6 & 12:32:40.32 & +12:46:22.53 & 22.089 & 21.187 & 9.7 & 0.78 & 2316 (6) & 12 \\
    GC 7 & 12:32:40.60 & +12:46:09.52 & 21.341 & 20.787 & 7.0 & 0.56 & 2286 (6) & 24 \\
    GC 8 & 12:32:40.89 & +12:46:23.02 & 23.041 & 22.616 & 7.2 & 0.58 & 2324 (10)$^\dagger$ & 4 \\
    GC 9 & 12:32:41.01 & +12:46:13.97 & 23.249 & 22.601 & 3.6 & 0.28 & 2319 (9)$^\dagger$ & 5 \\
    GC 10 & 12:32:39.06 & +12:46:04.71 & 21.376 & 20.801 & 27.9 & 2.23 & 2281 (7)$^\ddagger$ & 15 \\ 
    Star & 12:32:39.55 & +12:46:02.97 & 20.401 & 19.584 & 22.4 & 1.79 & 102 (6) & 18 \\ \hline
    VCC~9 GC1 & 12:09:22.27 & +13:59:32.42 & 22.010 & 21.445 & 0.5 & 0.037 & 1693 (8) & 14\\ \hline
    \end{tabular}
    \caption{A summary of the point sources we are able to analyse in our spectroscopic data. From left to right the columns are: 1) An object identifier, 2) Right Ascension, 3) Declination, 4) \textit{HST} F555W apparent magnitude, 5) \textit{HST} F814W apparent magnitude, 6) Projected distance to the VCC~1448 centre in arcsec, 7) Projected distance to VCC~1448 centre in kiloparsecs, 8) Recessional velocity with uncertainties in parentheses and 9) Signal to noise ratio of the spectrum analysed. $^*$ These GCs match those with measurements in \citet{Toloba2023}. $^\dagger$ These values were only measured in one of the two sky subtractions of the data and so should be considered less certain than other GC measurements. $^\ddagger$ This value is measured from the BH3/Large data as it lies out of the field of view of the BM/Medium data. The object labelled ``star" is likely a foreground star. The object labelled ``VCC~9~GC1" is a GC likely associated with VCC~9 which may be the galaxy's nucleus given its central positioning. All other objects are likely associated with VCC~1448. For VCC~9~GC1 columns 6/9 refer to the centre of VCC~9 and not VCC~1448. }
    \label{tab:gc_candidates}
\end{table*}

\begin{table}
    \centering
    \begin{tabular}{llll}
    \hline
    GC Sample & Mean GC $V_{\rm Sys}$ & $\sigma_{\rm GC}$ & $M_{\rm Dyn}$ \\
     & [\kms] & [\kms] & [$\times 10^{9} \mathrm{M}_{\odot}$] \\ \hline
    This work & 2304$^{+9}_{-9}$ & 27$^{+9}_{-6}$ & 7.49$^{+1.93}_{-1.17}$ \\
    T23 & 2310$^{+16}_{-17}$ & 48$^{+16}_{-11}$ & 8.16$^{+1.8}_{-0.99}$ \\
    This work + T23(<1$R_{\rm e}$) & 2298$^{+10}_{-10}$ & 29$^{+10}_{-7}$ & 2.88$^{+0.69}_{-0.39}$ \\
    This work + T23 & 2308$^{+9}_{-9}$ & 38$^{+9}_{-6}$ & 8.55$^{+1.73} _{-0.77}$ \\ \hline
    \end{tabular}
    \caption{The different mean GC systemic velocity ($V_{\rm Sys}$), GC velocity dispersion ($\sigma_{\rm GC}$) and tracer mass estimates ($M_{\rm Dyn}$) calculated in this work for VCC~1448. T23 refers the the GCs as measured by \citet{Toloba2023}. We present the results for both the combined sample and the combined sample for GCs within $1 R_{\rm e}$ only. All tracer mass estimates except that for the combined sample within 1 $R_{\rm e}$~(3.51 kpc) are measured within 6.48 kpc as the outermost GC is common to the two studies. When combining this study with T23 we preference our GC velocities for the two GCs we have in common. The addition of T23 to our data raises the velocity dispersion and the calculated tracer mass estimate.}
    \label{tab:GC_measurements}
\end{table}

\begin{figure}
    \centering
    \includegraphics[width = 0.47 \textwidth]{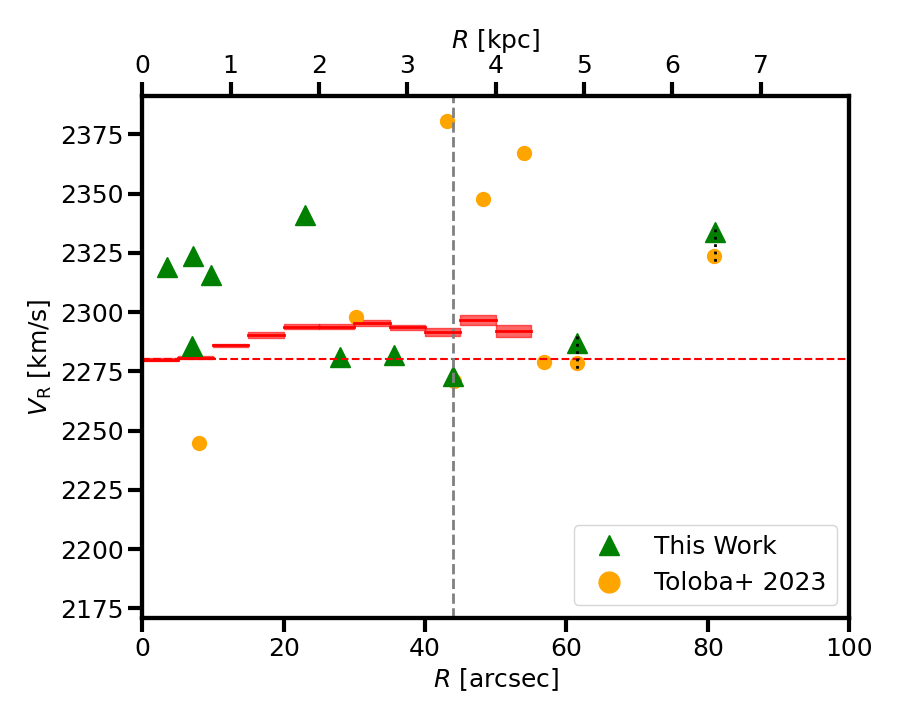}
    \caption{The phase space positions of the confirmed GCs around VCC~1448. GCs confirmed in this work are plotted as green triangles while GCs from \citet{Toloba2023} are included as orange circles. There are two GCs in common between the two studies that we join with a black dotted line. It is worth noting that there is also 1 GC in each study located near (x,y = 42, 2273) whose data points line on top of each other. These are not the same object. We include our stellar velocity curve for VCC~1448 as the red solid lines with 1-$\sigma$ uncertainty shading. The central recessional velocity of VCC~1448 is indicated by a dotted red line and the half-light radius of the galaxy is indicated by a dashed grey line. The majority of the confirmed GC candidates have recessional velocities above the galaxy's recessional velocity but this is at least partially due to the spatial coverage of the spectroscopic data.   }
    \label{fig:gc_phase_space}
\end{figure}

The KCWI data used in this work were observed on the night 2022, March 26 (N046, PI:Romanowsky). Conditions were clear with 1.1'' seeing. KCWI was configured using the medium slicer (16'' $\times$ 20'' field of view), BM grating and a central wavelength of 5075\AA. The spectral coverage of this configuration is $4650 - 5503$\AA. The instrumental resolution was measured from the arc calibration files and found to be R=5100 at 5075\AA~($\sigma_{\rm inst} = 24.9$~km s$^{-1}$). 

A total exposure time of 3600s was spent targeting a central pointing of VCC~9 (see Figure \ref{fig:vcc9_on_sky}) with 3600s on a nearby sky. Note that the sky position for VCC~9 is outside of the area plotted in Figure \ref{fig:vcc9_on_sky}. Three pointings were made targeting VCC~1448 along its major axis as indicated in Figure~\ref{fig:on_sky}. Total exposure times were  4800s, 3600s and 4800s for pointings 1, 2 and 3 respectively. In addition to these data a further 3600s of exposure were observed at each of the 2 sky pointings as indicated in Figure~\ref{fig:on_sky}. Both sky locations were chosen to include a compact source that was a candidate GC of VCC~1448.

We supplement these observations with two pointings of KCWI observed on the night 2021, April 16 as part of program Y228 (PI: van Dokkum). Conditions were clear with 0.7'' seeing. KCWI was configured using the large slicer (32'' $\times$ 20'' field of view), BH3 grating and a central wavelength of 5080\AA. The spectral coverage of this configuration is $4824-5315$\AA. Instrumental resolution in this configuration is approximately $\sigma_{\rm inst} = 25$~km s$^{-1}$ at 5000\AA. A total exposure time of 3600s was spent targeting each pointing of VCC~1448 as indicated by A and B in Figure~\ref{fig:on_sky}. The primary purpose of these additional KCWI observations was to recover the recessional velocities of GC candidates around VCC~1448. 

All data were reduced using the standard KCWI data reduction pipeline \citep{Morrissey}. Following the reduction, data cubes were trimmed to their good spatial and spectral wavelength ranges and then processed and sky subtracted as below. 

\subsection{VCC~9 Kinematic Measurements From Galaxy Light} \label{sec:vcc9_fitting}
We took the output standard star calibrated, non-sky subtracted `ocubes' of VCC~9 from the data reduction pipeline and collapsed them over both spatial directions into a single, non-sky-subtracted spectrum. At the distance of the Virgo cluster the KCWI medium slicer corresponds to a rectangular region of roughly 0.25 $R_{\rm e}$ within VCC~9 as seen in Figure \ref{fig:vcc9_on_sky}. A library of 9 sky spectra was then created using the offset sky observations around VCC~9 and VCC~1448 by collapsing the entire offset sky exposures over both spatial directions, excluding any compact sources in the data. This sky library, along with a template for galaxy emission, was input into the sky subtraction routine described in \citet{Gannon2020} to sky subtract our object data. Using this routine the data were modelled as a linear combination of our 9 sky principal components along with an old, metal poor template for galaxy emission from the library of \citet{Coelho2014}. After modelling, the sky portion of the model is subtracted to create a sky subtracted spectrum for each data cube. Barycentric corrections were then applied and each spectrum was median combined to produce the final science spectrum.

We used the code \texttt{pPXF} \citep{Cappellari2004, Cappellari2017} to measure a recessional velocity and velocity dispersion from the VCC~9 spectrum. Here we also followed the fitting method described in \citet{Gannon2020}. In brief, this involved fitting the sky-subtracted spectrum using the \citet{Coelho2014} synthetic stellar library in 241 different input parameter combinations to \texttt{pPXF}. This includes all combinations of 0-10th order additive and/or multiplicative polynomials along with 2 (i.e., pure Gaussian) and 4 (i.e., including $h_3$ and $h_4$) Gaussian moments. We do this to ensure that our final results are not dependent on our choice of input parameters to \texttt{pPXF}. We take our results as the median of the resulting parameter distributions with uncertainties as the 16th and 84th percentiles of these distributions. Using this process we measure a recessional velocity of 1690$\pm$0.4 \kms, a velocity dispersion of 27$\pm$0.6 \kms~and a spectral signal-to-noise ratio of 101\AA$^{-1}$ off of our spectrum. 

We note our uncertainties are drawn from the \texttt{pPXF} parameter distributions. There may be some concern that this method underestimates the true uncertainty in the results. In order to test this we took one of our spectra and created 100 random realisations of it based on uncertainties drawn from the residuals of its initial fit. These were then re-fitted with \texttt{pPXF}. The resulting parameter distributions for both recessional velocity and velocity dispersion are within the uncertainties derived by our primary fitting method.

These uncertainties however do not take into account any systematic uncertainty that may be present such as possible template mismatches between our fitting library and the data and any wavelength calibration issues. We estimate possible systematic errors based on the pixel sampling of our data as 10\% of the resolution of our data, i.e., $0.1\frac{c}{R} = 6$~\kms (see further \citealp{Robertson2017}). Further to this we find our ability to characterise the instrumental resolution of KCWI has a $\sim2$\% ($\sim0.5$~\kms) uncertainty which may affect our calculated velocity dispersions. We incorporate these uncertainties in quadrature to the relevant measurements for the remainder of the paper.

Our recessional velocity is within the uncertainties of those previously listed for VCC~9 (e.g., SDSS; \citealp{SDSS2009}). Furthermore, this recessional velocity and velocity dispersion are in reasonable agreement with those listed from William Herschel Telescope spectroscopy of VCC~9 by \citet[1674$\pm$6~\kms; 26$\pm$4.6 km s$^{-1}$]{Toloba2014}. We suggest the good agreement of our velocity dispersion to that from \citet{Toloba2014} allows us to adopt their rotation value for the remainder of the paper as we do not sample out to the half-light radius where we require this measurement for comparison in Section \ref{sec:discussion}. The fitting process used to produce our value includes the higher order Gaussian moments $h_3$ and $h_4$ in half of the fits to ensure their inclusion does not drastically affect our results. These additional parameters do not significantly change the recessional velocity/velocity dispersion from those fits that do not include them. We display the resulting spectrum for VCC~9, along with an example best fitting \texttt{pPXF} fit, in Figure \ref{fig:spectra} (top panel). 

\subsection{VCC~1448 Kinematic Profile Measurements From Galaxy Light} \label{sec:vcc1448_profile}

We performed two separate extractions of spectra to create a radial profile for VCC~1448. In the first, we created a crude radial profile for VCC~1448 by spatially collapsing each of our three KCWI pointings along the spatial axes into a single spectrum. These are then sky subtracted, median combined and fitted using \texttt{pPXF} as described for VCC~9 in subsection \ref{sec:vcc9_fitting}. In each case, compact sources in the field of view are masked prior to spectral extraction. We display these three spectra,  along with an example best fitting \texttt{pPXF} fit, in Figure \ref{fig:spectra}. 

For our second extraction of a radial profile we desired higher spatial resolution (i.e., more radial bins). During early attempts to extract spectra in elliptical apertures on VCC~1448 we discovered the WCS coordinates available in KCWI reduced data cubes did not have the required accuracy for cube stacking. To correct for this inaccuracy we took the brightest point source in each data cube, fitted a Gaussian profile, measured its centre and reassigned the fits header WCS to this position.

We took the centre of VCC~1448 to be at RA $=$ 188.170009 deg. and Dec $=$ +12.771085 deg. (both J2000), the galaxy's position angle to be 88 degrees and its ellipticity $\epsilon=0.19$ \citep{Ferrarese2020}. We then used the galaxy centre, position angle and axis ratio to place elliptical apertures on the KCWI data using the \texttt{photutils} routine \texttt{SkyEllipticalAperture} \citep{Bradley2023}. The initial aperture had a semi-major radius of 3'' with the 4 subsequent annuli stepping out by 3'' in radius in each aperture. Following these initial 5 apertures, 8 more were added that each stepped out in 5'' in radius each aperture. Apertures were limited to one side of the minor axis of the galaxy to ensure that they did not sample both the positive and negative side of the galaxy's rotation. Our apertures for extraction can be seen on sky in Figure \ref{fig:on_sky}. At our assumed 16.5 Mpc distance for Virgo our full profile reaches out to 4.4 kpc which corresponds to $\sim$ 1.25 $R_{\rm e}$.

For each radial bin, spectra are extracted then sky subtracted and fitted using \texttt{pPXF}, as described for VCC~9 in subsection \ref{sec:vcc9_fitting}. Compact sources are masked in each aperture before the extraction of a spectrum. The final recessional velocity, velocity dispersion and signal-to-noise for each radial bin are visible in Figure \ref{fig:radial}. Measured values of recessional velocity and velocity dispersion for our first, crude extraction are in good agreement with those measured in similar radial regions of our second extraction. For the remainder of this paper, we will use velocity dispersions and recessional velocities from the second extraction. 

We note that while this extraction steps out along one side of the galaxy, we also sample out to small radii along the major axis in the other direction. We have also derived the results for the first 4 apertures available to step out in this direction. They are within the uncertainties of the results displayed in Figure \ref{fig:radial} for both velocity dispersion and rotation. However, for these data, the rotation was blue-shifted with respect to the systemic velocity of the galaxy along the line of sight. 

We measure the barycentric corrected recessional velocity of VCC~1448 to be 2280$\pm$6 \kms~(i.e., the value from the central bin of our second radial extraction). This value is within $2-\sigma$~uncertainties of the systemic velocity of the GC system reported by \citet[$2310^{+16}_{-17}$ \kms]{Toloba2023}. Our spectroscopic recessional velocity amends the reporting of VCC~1448 being at 2572$\pm$15 \kms~ by \citet{Binggeli1985}. We note the \citet{Binggeli1985} measurement has been repeated (with small alterations) in various catalogues listing VCC~1448, however no other unique recessional velocity measurement exists for the galaxy that we are aware of. Given the high S/N of our various extracted spectra, their derived redshift seems unlikely. We suggest the cause of this difference is that their listed recessional velocity is from the HI thought to be associated with the galaxy \citep{Huchtmeier1985}. It is instead likely this gas is no longer associated with VCC~1448, although it is possible it is in the process of being ram pressure stripped \citep{Vigroux1986}. We use our recessional velocity when calculating projected rotation in the galaxy subject to additional corrections below. 

\subsection{Azimuthal Correction}
We note that in coadding our spaxels we will be adding spaxels that do not align with the semi-major axis of VCC~1448 along which rotation is usually quoted. To correct for this issue we derive a correction for each annulus using a formula derived from equation 3 of \citet{Foster2011}. This formula requires the systemic velocity of the galaxy ($V_{\rm Sys}$), the observed recessional velocity ($V_{\rm Obs}$), the position angle of the observed spaxel ($PA_i$), the position angle of the kinematic axis ($PA_{\rm kin = phot}$), the kinematic axis ratio ($q_{\rm kin=phot}$) and the position angle of the upper limit of the wedge of annulus observed ($\theta_{\rm max}$). For both the kinematic position angle and axis ratio we assume it to be equivalent to the photometric axis ratios. The equation to calculate the rotational velocity ($ V_{\rm Rot}$) then takes the form:

\begin{equation} \label{eqtn:azimuthal}
    V_{\rm Rot} = (V_{\rm Obs} - V_{\rm Sys}) \times \int_{PA_{\rm kin = phot}}^{\theta_{\rm max}} 2 \times \sqrt{1+ \left( \frac{\tan \left( \theta - PA_{\rm kin = phot}\right)}{q_{\rm kin=phot}}\right)^{2}} d\theta
\end{equation}

where we calculate the maximum position angle in our wedge to evaluate it over as:

\begin{equation}
    \theta_{\rm max} = PA_{\rm kin = phot} + \arctan(\frac{0.64 {\rm kpc}}{(\frac{R_{\rm min}+ R_{\rm max}}{2}) {\rm kpc}})
\end{equation}



which is derived by the geometry of our KCWI pointings at the distance of VCC~1448. In practice equation \ref{eqtn:azimuthal} has little effect on our measured rotation at large radii, with most of the correction being evident in the central 4-5 apertures ($\sim$ 0.33 $R_{\rm e}$). 

\subsection{Inclination corrections}
In addition to our azimuthal correction, it is worth noting that VCC~1448 is not obviously edge-on. As such, any rotation measured is likely only a lower limit to the true rotation in the galaxy. Under the assumption of a thin disk, a lower limit on the inclination angle (0 degrees = face-on; 90 degrees = edge-on) may be calculated from the photometric ellipticity using $i = \cos^{-1}(1-\epsilon)$. For VCC~1448 this equates to $i=36$~degrees. Using the standard $\sin{i}$~inclination correction this implies that an assumption of viewing VCC~1448 edge on (i.e., $V_{\rm Measured} = V_{\rm Intrinsic}$) may underestimate the true rotation of the galaxy by a factor of up to 1.7$\times$. Therefore, we will also include the results of the intrinsic rotation of VCC~1448 if we are observing it at inclination angles of 75, 60, 45 and 36 degrees (i.e., the lower limit) when assuming VCC~1448 is viewed edge-on throughout important sections of the remainder of the paper. In practice, none of these inclination corrections are large enough to alter the conclusions that will be drawn. 

After the inclusion of the inclination and azimuthal corrections, there is a flattening of the rotation curve in VCC~1448 at about 0.5 $R_{\rm e}$ (see Fig. \ref{fig:radial}). However, the maximum rotational velocity in a galaxy is not usually reached until beyond 1 $R_{\rm e}$~\citep{Beasley2009, Geha2010}. It is therefore possible that we have not yet reached the maximum rotational velocity of VCC~1448. For the remainder of the paper, we therefore compare rotation as measured within the half-light radius of each galaxy rather than the maximum rotation of each galaxy.

\subsection{Dynamical Masses} \label{sec:dynamical_masses}
Now that we have a velocity dispersion and rotational velocity we use them to calculate a mass using the formula of \citet{wolf2010}. This formula uses the 2-D circularised half-light radius $R_{\rm e, Circ}$ and the line of sight velocity dispersion within that radius ($\sigma$) and takes the form:

\begin{equation} \label{eqtn:wolf}
    M(<\frac{4}{3} R_{\rm e, Circ}) = 930 \left( \frac{\sigma^2}{\rm km s^{-1}} \right) \frac{R_{\rm e, Circ}}{\rm pc}
\end{equation}

We note however that the formula of \citet{wolf2010} was designed for pressure-supported systems and that both VCC~1448 and VCC~9 have significant rotation support within the half-light radius. To account for this rotation we instead use the light-weighted average $V_{\rm rms} = \sqrt{\langle V \rangle^{2}_{\rm Rot} + \langle \sigma \rangle^{2}}$ within 1 $R_{\rm e}$ as the ``$\sigma$" in Equation \ref{eqtn:wolf} \citep{Dabringhausen2023}. Using this method we derive a dynamical mass of 1.96$\pm$0.68$\times$10$^9$~M$_{\odot}$ within 4.67 kpc for VCC~1448. It is worth noting that this total dynamical mass is only slightly greater than the expected stellar mass within the half-light radius ($\sim 1.3\times 10^{9}$~M$_\odot$) and is within 1-$\sigma$ uncertainties of it. Using the same process for VCC~9 we derive a dynamical mass of 2.69$\pm$2.29$\times$10$^9$~M$_{\odot}$ within 3.96 kpc. 


Finally, our rotation and velocity dispersion allows us to investigate the stellar motions in the galaxy. We determine the flux-weighted average rotational velocity over velocity dispersion ($\langle V \rangle / \langle \sigma \rangle$) based on the commonly used formula of \citet{Cappellari2007}:

\begin{equation} \label{V/sigma, equation}
    \frac{\langle V \rangle}{\langle \sigma \rangle} = \sqrt{\frac{\langle V^{2} \rangle}{\langle \sigma^{2} \rangle}} = \sqrt{\frac{\sum_{n=1}^{N} F_{n} V^{2}_{n}}{\sum_{n=1}^{N} F_{n} \sigma^{2}_{n}}}
\end{equation}

with $F_{n}$, $\sigma_{n}$~and $V_{n}$~as the measured flux, velocity dispersion and rotational velocity in each bin respectively. 

We apply this formula to the central bins of our profile (i.e., those within 1 $R_{\rm e}$) where we calculate the flux values for each bin based on the measured S\'ersic profile of the galaxy. Our final $\frac{\langle V \rangle}{\langle \sigma \rangle}$ value for VCC~1448 is 0.45. We report this value along with the flux-weighted average velocity dispersion ($\langle \sigma_{\rm e} \rangle$) and the flux-weighted average rotational velocity corrected to the major axis ($\langle V_{\rm rot, e} \rangle$) in Table \ref{tab:target_Summary}. 

\subsection{GC Candidate Recessional Velocities}
In addition to the BM, Medium, 5075\AA~central wavelength data (hereafter BM/M) we use for kinematics we also have access to BH3, Large, 5080\AA~central wavelength KCWI data (hereafter BH3/L) in two pointings as described in Section \ref{sec:kcwi_data}. The positioning of these two pointings is visible in Figure \ref{fig:on_sky}. We reduced and stacked these data in the same manner as the BM/Medium data but here we performed an additional flat fielding correction prior to stacking as described in \citet{Gannon2020} to correct for a low-level gradient observed in the images. For the BM/Medium data we do not perform this correction as it is not possible to disentangle a gradient caused by the galaxy in the background from this flat fielding error. A peculiarity of KCWI is its rectangular spaxels. In the case of this Large slicer data, the sampling is such that the spatial resolution is pixel-limited along the long axis of the slicer while being seeing-limited on the short axis. After rebinning and stacking, this causes point sources such as GCs to appear oblong in their shape.

We extracted spectra for all visually identified compact sources in the stacked data cubes in two manners: 1) by placing 1'' diameter apertures on each source and subtracting off a nearby region of the data cube as the background (sky + galaxy); 2) by placing a 1'' diameter aperture on each source and subtracting off an annulus with inner diameter 1.5'' and outer diameter 3'' as background (sky + galaxy). In the case of the BM/M data these apertures were circular. In the case of the BH3/L data these apertures were oblong along the long axis of the slicer to account for the different spatial resolution in that direction. In total, we extracted spectra for 35 objects from both VCC~1448 and VCC~9, 11 of which were duplicated between two KCWI configurations. In general, the first extraction, which uses a nearby offset sky region for background subtraction proved better able to recover a recessional velocity than using an annular sky background. For the remainder if this paper we report values from this extraction. 

We fitted all of these spectra with \texttt{pPXF} using 2 Gaussian moments (i.e., no $h_3$~or $h_4$) along with 6 additive and multiplicative polynomials. We ensure our final reported redshifts are robust to the initial redshift estimate by running \texttt{pPXF} with 51 different input recessional velocities linearly distributed from 200 \kms~above to 200 \kms~below the recessional velocity of the respective galaxy host. Redshifts were only reported when there is a clear independence of recovered GC velocity on initial \texttt{pPXF} guess. 10 GCs satisfy this criteria for VCC~1448 while only 1 satisfied this criteria for VCC~9. It is possible the confirmed GC for VCC~9 is the nucleus of the galaxy given its central location. 1 object was identified as a foreground star. For the 11 objects that were duplicated in both KCWI configurations 3 had recessional velocities in agreement between the configurations and 2 had recessional velocities that could only be deemed reliable in the BM/Medium grating. For all other GC candidates remaining in either KCWI configuration, we were unable to measure a recessional velocity due to low S/N. We present the GC position, magnitude, projected distance to VCC~1448/VCC~9 centre and recessional velocity in each configuration in Table \ref{tab:gc_candidates}. The recessional velocities have been barycentric corrected \citep{TollerudBary}. Magnitudes are quoted from the automatic \texttt{DAOPhot} Hubble Legacy Archive catalogues generated for the \textit{Hubble Space telescope} (\textit{HST}) imaging from \citet{Miller1998} and \citet{Seth2004}.

In Figure \ref{fig:gcs_onsky} we display the position of our GCs around VCC~1448 and the foreground star in the sky with colour coding corresponding to their absolute velocities around the recessional velocity of VCC~1448. The majority of our GCs reside at a higher recessional velocity than the galaxy. There is no obvious sign of rotation in the GC system.

In Figure \ref{fig:gc_phase_space} we then place our GCs in the phase space centred on VCC~1448. We include GC positions from the galaxy centre but note that our rotation curve for the galaxy which lies along the semi-major axis. Formally this does not take into account the ellipticity of VCC~1448, where projected GC distances away from the semi-major axis are ``further out" in the galaxy (i.e., a larger fraction of the galaxy radius along that axis) for a similar observed distance. We do not apply these corrections so as to include and be comparable to the data plotted in \citet{Toloba2023} figure 4. Two of our GCs are in common. For our GC1 \citet{Toloba2023} measured a recessional velocity of 2324$\pm$3 \kms and for our GC3 they measure 2279$\pm$8 \kms. Both values are slightly lower than our measurements but within $2-\sigma$~joint uncertainties. Again, we can see that the GCs, measured both in this work and in \citet{Toloba2023}, have higher recessional velocities than the galaxy on average. We note that this is at least partially due to the sampling of our KCWI data being asymmetric along only one side of the galaxy. 


We feed our 10 GC recessional velocities into the GC velocity dispersion fitting code outlined in Haacke et al. (in prep). Following appendix A in \cite{Doppel21}, the code uses the python package \texttt{emcee} \citep{emcee} to fit the mean velocity of the GC system and the velocity dispersion for the GCs using Monte Carlo Markov Chain exploration of the standard Gaussian maximum likelihood equation as used in e.g., \citet{vanDokkum2018} or \citet{Muller2020}. Individual GC velocities are weighted by their uncertainties when included in this fitting process. A flat prior is used in the range 0-100 km s$^{-1}$. Final values are reported from the median of the posterior distribution with 16th and 84th percentiles as the uncertainties. From this analysis, and using only our 10 GC measurements we report a GC-system velocity dispersion of 27$^{+9}_{-6}$~km s$^{-1}$, which is in good agreement with our stellar velocity dispersion of 24.0$\pm$1.6\kms. Interestingly our fitting also revealed a mean GC velocity of 2304$^{+9}_{-9}$~km s$^{-1}$, $\sim 20$~km s$^{-1}$~more than that we measure for the galaxy.


We note that our GC velocity dispersion is well below the 48$^{+16}_{-11}$~km s$^{-1}$~reported by \citet{Toloba2023} based on their sample of 9 GCs. However our average GC-system velocity is in good agreement with their mean velocity, 2310$^{+16}_{-17}$~km s$^{-1}$. Using their data (E Toloba, priv. comm.) our dispersion fitting code was able to produce results in agreement with their published values. We therefore combine both our data with those from \citet{Toloba2023}, using our recessional velocities for GCs that appear in both studies. For the combined sample we derive an average GC-system velocity of 2308$^{+9}_{-9}$~km s$^{-1}$ with a dispersion of 38$^{+9}_{-6}$~km s$^{-1}$. As expected this value is between those recovered with either survey used independently.

In order to ensure our measured velocity dispersions were not being overly driven by an outlier GC we iteratively re-fitted the velocity dispersion for these samples each time excluding 1 GC measurement. For the GC sample coming from this work the maximum difference in the velocity dispersion from the fiducial value was $\sim$4 \kms. Likewise for the combined GC sample of our work and \citet{Toloba2023}, the maximum difference in the velocity dispersion was also $\sim$4 \kms. In both cases, the exclusion of the GC that drove the maximum difference resulted in lower velocity dispersions. As both differences are well within the uncertainties of our fiducial values, we are satisfied that there does not exist an outlier GC overly affecting our results. We summarise these values for VCC~1448 in Table \ref{tab:GC_measurements}.


 In general, the mean of the GC system is expected to be representative of that of the galaxy (see \citealp{Forbes2017}, fig. 1). Here however we measure a mean velocity of the GC system that is larger than that we measure for the galaxy. Fixing the GC mean to be that of the galaxy would cause a larger velocity dispersion to be measured from our data. However, this offset is at least partially due to our uneven sampling of the Gaussian distribution of GC velocities as the result of the uneven spatial coverage of our KCWI pointings. This reinforces the benefits of our study also targeting VCC~1448's stellar body to get an independent measurement of its recessional velocity.

However, there also exists a difference between the GC velocity dispersion measured from our sample and that of \citet{Toloba2023}. \citet{Toloba2023}'s GC sample comes from GCs that are on average further from the centre of VCC~1448 than our data (see further Figure \ref{fig:gc_phase_space}). If VCC~1448 exhibits a strongly rising velocity dispersion profile in its GC system it may be expected to see such a difference between the velocity dispersions measured from our GCs and those of \citet{Toloba2023}. A rising velocity dispersion profile is not seen in the stellar profile however, which is approximately flat with radius (see further Figure \ref{fig:radial}).



Finally, we use the combined GC projected positions and recessional velocities to estimate a mass for the galaxy using a Tracer Mass Estimator \citep{Watkins2010}. Here we calculate a mass within the galactocentric radius enclosing the outermost GC ($r_{\rm out}=6.48$~kpc), by inputting the velocities of the tracers along the line of sight ($v_{\rm los}$) and their radii ($r$). When calculating the tracer velocities we subtract off VCC~1448's systemic velocity as listed in Table \ref{tab:target_Summary}. We also need to make assumptions on the log-slope of the gravitational potential ($\alpha$), the orbital (an)isotropy ($\beta$) and the power-law slope of the GC density profile ($\gamma$). The tracer mass estimator then takes the form:

\begin{equation}
    M \left(< r_{\rm out}\right) = \frac{C}{G} \langle v_{\rm los} ^{2} r^{\alpha}\rangle
\end{equation}
with
\begin{equation}
    C = \frac{\alpha + \gamma + 1 - 2 \beta}{I_{\alpha, \beta}} r_{\rm out}^{\left(1-\alpha\right)}
\end{equation}
and
\begin{equation}
    I_{\alpha,\beta} = \frac{\pi^{\frac{1}{2}} \Gamma\left(\frac{\alpha}{2}+1\right)}{4 \Gamma\left(\frac{\alpha}{2} + \frac{5}{2}\right)} \left(\alpha +3-\beta\left(\alpha+2\right)\right)
\end{equation}

where $\Gamma$ refers to the gamma function. 

Here we assume $\alpha=-0.7$, $\beta=0$ and $\gamma$=1.25. This is equivalent to assuming a combined NFW \citep{NFW} and stellar mass potential along with completely isotropic orbits. Our GC density slope $\gamma$=1.25 assumption is based on reported GC density profiles of GC-rich Virgo dwarfs of similar magnitude in \citet{Beasley2016}. Based on these assumptions, and the combined 17 GC sample from our study with \citet{Toloba2023}, we measure an enclosed mass of 8.55$^{+1.07} _{-0.48}$$\times$10$^9$~M$_\odot$ within 6.48 kpc for VCC~1448. Enclosed masses based on just the GCs in this work along with just the \citet{Toloba2023} GC sample are listed in Table \ref{tab:GC_measurements}.


Uncertainties for our enclosed mass were estimated via the 16th and 84th percentiles of 1000 mass estimations in which each GC recessional velocity had been perturbed by a random number drawn from a Gaussian of standard deviation equivalent to the quoted uncertainty. We note however that our mass uncertainty does not include the systematic error caused by our assumptions on $\alpha$, $\beta$~and $\gamma$. A $\pm$0.3 change in $\alpha$ changes the measured enclosed mass by at most 12\%. A $\pm$0.1 change in $\beta$ will only change our measured mass by $\sim\pm$6\%. A $\pm$0.25 change in our assumed $\gamma$ will change our measured mass by $\sim\pm$15\%.

\begin{figure}
    \centering
    \includegraphics[width = 0.47 \textwidth]{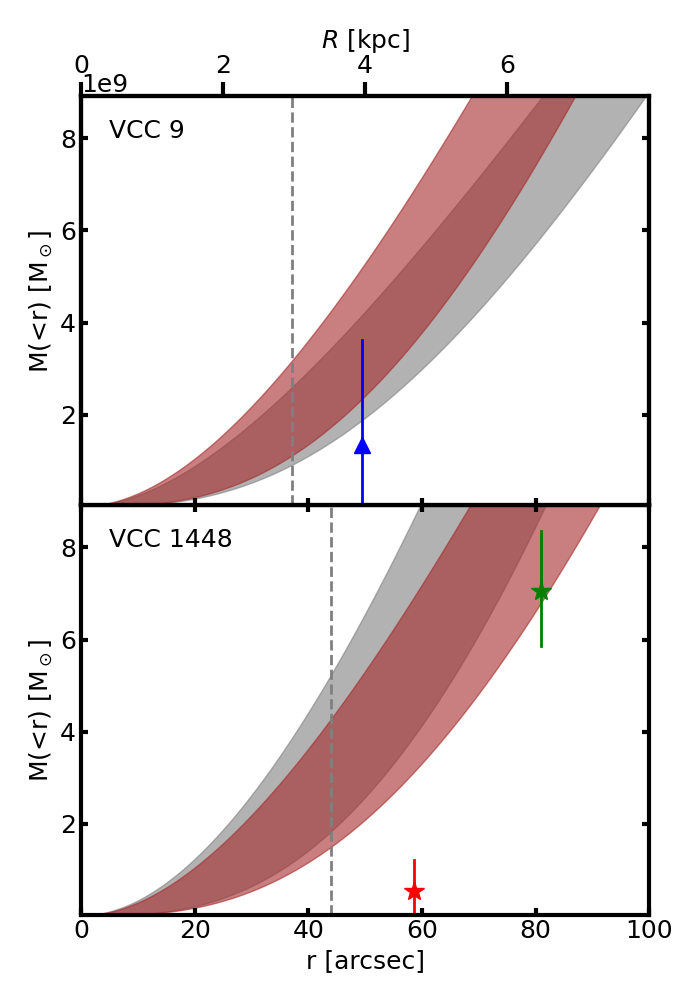}
    \caption{Enclosed mass \textit{vs.} radius. \textit{Upper:} An enclosed dynamical mass based on VCC~9's stellar motions is plotted as a blue triangle. \textit{Lower:} An enclosed dynamical mass for VCC~1448 based on its stellar motions (i.e., $V_{\rm rms})$ is plotted as a red star and an enclosed dynamical mass for VCC~1448 based on its GC motions is plotted as a green star. For both, the projected half-light radius for each galaxy is marked with a dashed grey vertical line. The grey-shaded region corresponds to dark matter halos with total halo mass based on the prediction from the number of GCs in its GC system. The brown shaded region is instead the total halo mass prediction from the \citet{Danieli2023} stellar mass -- halo mass relation for a galaxy of VCC~9's / VCC~1448's stellar mass. For both regions we plot halos ranging from cuspy NFW halos to completely cored dark matter halos with core radius equal to 2.75$\times R_{\rm e}$. VCC~9's dynamical mass measurement agrees with both the expectation of the stellar mass -- halo mass and the GC number -- halo mass relationship. VCC 1448’s dynamical masses are in better agreement with the expected total halo mass derived from the stellar mass -- halo mass relationship, rather than the GC number -- halo mass relationship.}
    \label{fig:mass_profile}
\end{figure}

\begin{figure}
    \centering
    \includegraphics[width = 0.48 \textwidth]{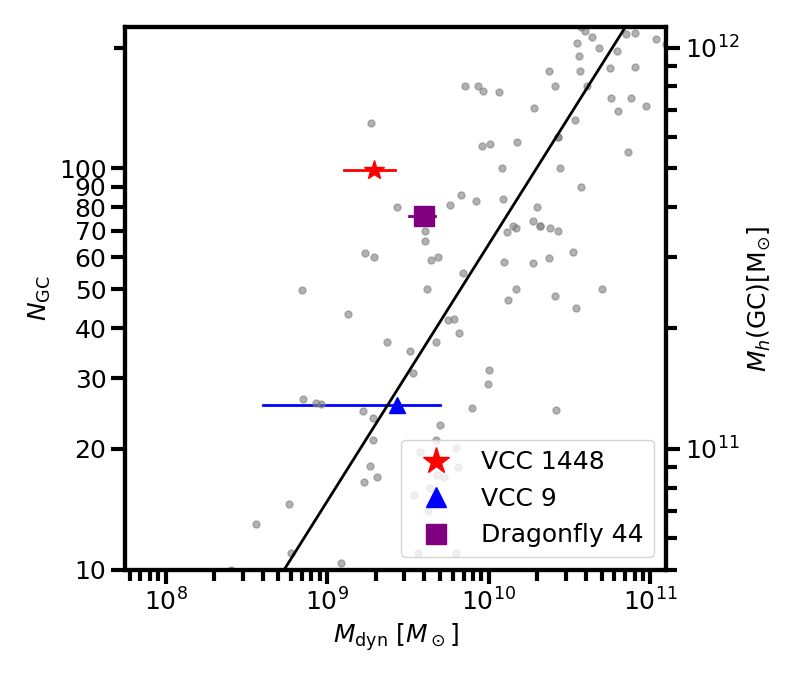}
    \caption{GC number \textit{vs.} dynamical mass within the half-light radius. The right hand axis shows the total halo mass expected given the GC number -- halo mass relationship of \citet{Burkert2020}. VCC~9 is plotted as a blue triangle. Dragonfly 44 is plotted as a purple square with kinematic data from \citet{vanDokkum2019b}. VCC~1448 is plotted as the red star. Small grey symbols show previously studied galaxies from the catalogue of \citet{Harris2013} with a solid line showing a fit to the low mass galaxies in the catalogue. Both VCC~1448 and Dragonfly~44 have large numbers of GCs for their dynamical mass while VCC~9 agrees with the galaxies from the \citet{Harris2013} catalogue. }
    \label{fig:Ngc_mdyn}
\end{figure}

\begin{figure}
    \centering
    \includegraphics[width = 0.48 \textwidth]{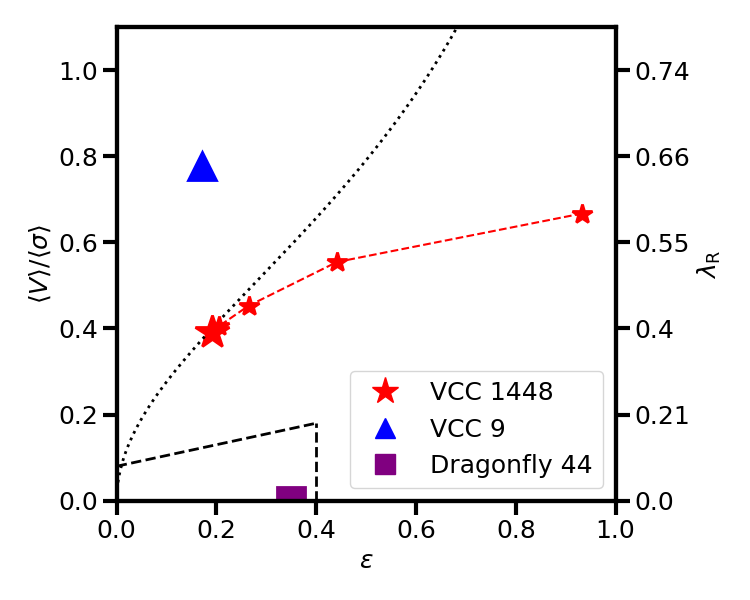}
\caption{$\frac{\langle V \rangle}{\langle \sigma \rangle}$ \textit{vs.} ellipticity ($\epsilon$). We include a second y-axis where we convert $\frac{\langle V \rangle}{\langle \sigma \rangle}$ into the commonly used $\lambda_{R}$ using \citet{Penoyre2017}. VCC~1448 is the large red star, VCC~9 is the blue triangle and Dragonfly~44 is the purple square. The black dotted line corresponds to an isotropic orbital structure within this parameter space \citep{GalacticDynamics, Cappellari2007}. The black dashed line denotes the region defining the separation between fast and slow central rotators as proposed in \citet{Emsellem2011} and updated by \citet{Cappellari2016}. We show the possible intrinsic positioning of VCC~1448 given inclinations of 75, 60, 45 and 36 degrees as the connected smaller red stars. VCC~1448 is a noticeably more dispersion-supported system than VCC~9. Both are classified as fast rotators using the well-established convention for elliptical galaxies. In contrast to VCC~9 and VCC~1448, Dragonfly~44 is a slow rotator, being almost completely dispersion-dominated. The three galaxies lie in very different regions of the parameter space. }
    \label{fig:Vsigmavsellipticity}
\end{figure}

\begin{figure}
    \centering
    \includegraphics[width = 0.48 \textwidth]{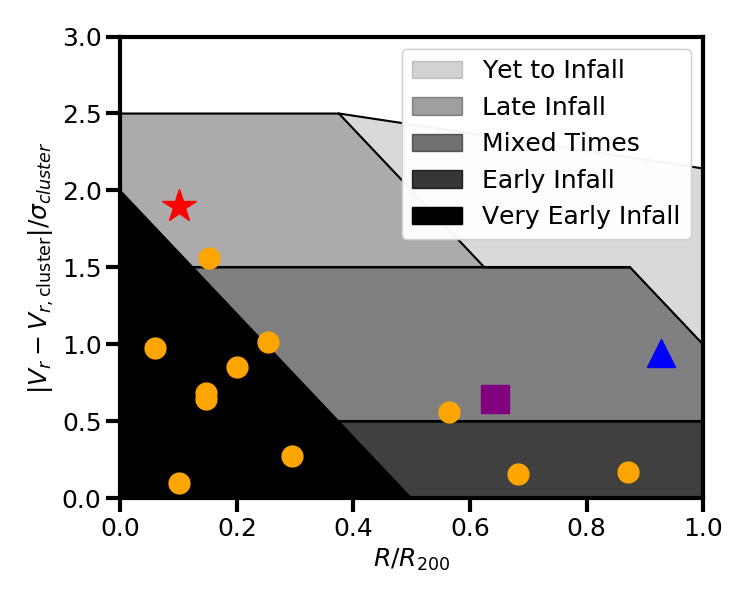}
    \caption{Galaxy radial velocity minus the galaxy cluster mean velocity normalised by the cluster's velocity dispersion \textit{vs.} normalised radius from cluster centre. Shaded regions are from the cosmological simulations of \citet{Rhee2017}, whereby darker regions correspond to galaxies with predominantly earlier infall time. This is as indicated in the legend. VCC~1448 is plotted as a red star, VCC~9 is plotted as a blue triangle and Dragonfly~44 is plotted as a purple square. Dwarfs in the Virgo cluster from the study of \citet{Liu2016} are shown as orange circles. The GC-rich galaxies Dragonfly~44 and VCC~1448 do not follow this picture, with both likely falling in at later times.}
    \label{fig:rhee_diagram}
\end{figure}

\section{Discussion} \label{sec:discussion}
\subsection{Mass Measurement Comparison to Mass Profiles}

We are now equipped with a single mass measurement for VCC~9 and two mass measurements for VCC~1448 at differing radii. In Figure \ref{fig:mass_profile} we compare these mass measurements halo profiles based on total halo masses from the GC number and stellar mass -- halo mass relationships (see below).

We plot our enclosed dynamical mass measurements using both the \citet{wolf2010} formula (for both VCC~9 and VCC~1448) and the \citet{Watkins2010} tracer mass estimation formula (for VCC~1448). All mass measurements have the stellar mass within their radius subtracted using \citet{Price2022} to leave solely the dark matter mass. Finally, we include for comparison the region of halo profiles expected with total mass: 1) based on the expectation from its large GC count and the GC number -- halo mass relationship (VCC~9~$M_{\rm Halo}=$1.29$\times10^{11}$~M$_{\odot}$; VCC~1448~$M_{\rm Halo}=$4.97$\times10^{11}$~M$_{\odot}$; \citealp{Burkert2020}); and 2) based on the expectation given VCC~9's/VCC~1448's stellar mass and the stellar mass -- halo mass relationship (VCC~9~$M_{\rm Halo}=$2.47$\times10^{11}$~M$_{\odot}$; VCC~1448~$M_{\rm Halo}=$2.51$\times10^{11}$~M$_{\odot}$; \citealp{Danieli2023}). The halo mass profiles are plotted assuming limiting cases of a cuspy NFW \citep{NFW} and a \citet{Read2016} cored profile with core radii set to be 2.75$\times R_{\rm e}$. These correspond to minimal and maximal core formation in the halo, and thus minimal and maximal central masses are expected within the central regions of the dark matter halo. In each case, we take the halo concentration from \citet{Dutton2014}. 

For VCC~9 we find our mass measurement is in good agreement with the expected mass for dark matter halos of total mass predicted by either the stellar mass -- halo mass or GC number -- halo mass relationship. For VCC 1448 our crude mass profile based on two mass measurements indicates that it likely does not reside in the massive dark matter halo expected given the reported rich GC system for the galaxy or the stellar mass – halo mass relationship.  For VCC~1448 our crude mass profile there is a better agreement between our mass measurements and the expected dark matter halo mass based on the stellar mass – halo mass relationship than the GC number halo mass relationship. We note Figure \ref{fig:mass_profile} does not include the scatter in the \citet{Burkert2020} relationship. The addition of this $\sim$0.3 dex scatter would place our mass measurements within 1-$\sigma$ uncertainties.


However there is still some tension with our stellar dynamical mass measurement for VCC~1448. In order for this measurement to fit with the massive dark matter halo expected given either relationship, our choice of a normal concentration parameter ($c=8.93$) for VCC~1448 must be incorrect. Recent findings by \citet{Toloba2023} showed higher than usual concentration parameters are needed to explain their studied Virgo dwarf galaxies' dynamical masses. However, VCC~1448 would require the opposite. Namely, here a lower concentration parameter would be required to reproduce our measured dynamical masses. We take this as unlikely given the cored halo mass profiles are not in good agreement with our data. Strictly speaking, although we generate \citet{Read2016} mass profiles using a standard halo concentration, the mass profile modifications to create the core lower the concentration of the resulting mass profile. In this way, our choice to plot cored halos has a similar effect to lowering the concentration of the dark matter halo. A further lowering of the concentration of VCC~1448's dark matter halo to fit the expected total halo mass of the GC number -- halo mass relationship is therefore unlikely.


It is worth noting that while many suggest that VCC~1448 has smooth isophotes (see e.g., \citealp{Reaves1962, Knezek1999}), some have suggested the presence of a central bar and/or knots and/or shells in its isophotes (see e.g., \citealp{Reaves1983, Vigroux1986, Knezek1999, Lim2020}). These may indicate a recent interaction/merger in the galaxy's formation history, however, we note many of the knots appear to be candidate clusters in higher resolution imaging \citep{Seth2004}. Alternatively, it has been suggested that VCC~1448 may currently be in the transition from a late-type to an early-type dwarf galaxy driven by ram pressure stripping in the Virgo Cluster (see e.g., \citealp{Knezek1999}). If either is the case, VCC~1448's dark matter halo may not currently be in equilibrium and the simplified comparisons such as that we make in Figure \ref{fig:mass_profile} may not be appropriate. Performing more extensive Jeans model-based fitting to our data for VCC~1448 to better explore its dark matter halo will be the subject of future work.


\subsection{Dynamical Masses versus Globular Cluster System Richness}
It has been observed that the GC-richness of a galaxy correlates with its dynamical mass within the half-light radius over many orders of magnitude for normal galaxies \citep{Harris2013}. Given the differing GC counts of VCC~9 and VCC~1448, and thus differing expected halo masses, there is an expectation that VCC~1448 should be more massive, possibly following this trend. 

To investigate this we have produced Figure \ref{fig:Ngc_mdyn}. Here we plot the number of globular clusters ($N_{\rm GC}$) \textit{vs.} the dynamical mass within the half-light radius ($M_{\rm dyn}$) for VCC~9, VCC~1448 and Dragonfly~44. Similar to \citet{Forbes2024}, we include the \citet{Harris2013} galaxy sample along with a line of best fit to galaxies in the sample with $\log(M_{\rm dyn} / M_{\odot})<11$. While VCC~9 follows the sample of normal galaxies from \citet{Harris2013}, VCC~1448 and Dragonfly~44 both exhibit greater GC richness than other galaxies of their dynamical mass.

When considering the positioning of VCC~1448 in this space, inclination effects may cause us to underestimate the intrinsic rotation of the galaxy, and thus underestimate the dynamical mass. Formally, our choice to plot the measured dynamical mass of VCC~1448 in Figure \ref{fig:Ngc_mdyn} is equivalent to assuming our galaxy is being viewed edge-on, which may not be accurate given its low ellipticity ($\epsilon=0.19$). A galaxy with higher intrinsic ellipticity viewed more face-on would also result in our rotation measurement while the intrinsic rotation would be higher \citep{Cappellari2016}, increasing the dynamical mass. In practice, this effect can only lead to a modest increase in dynamical mass which is insufficient to place VCC~1448 on the relation for normal galaxies we show in Figure \ref{fig:Ngc_mdyn}. 

It is also worth considering the type of galaxies that both Dragonfly~44 and VCC~1448 are being compared to in Figure \ref{fig:Ngc_mdyn}. Due to their elevated GC richness for dwarfs of a similar stellar mass, the galaxies with similar GC richness in the panel tend to be of much higher stellar mass, which is included in the dynamical mass measurement for each galaxy. If this stellar mass component is removed from the dynamical mass so that the remaining mass is predominantly dark matter there is expected to be better agreement between Dragonfly~44/VCC~1448 and the normal galaxies. In this way, the GC number -- dynamical mass relationship may be seen as an extension of the GC number -- halo mass relationship. A separate effect that also goes to place Dragonfly~44 and VCC~1448 on the relationship is that of dark matter halo profiles. Both VCC~1448 and Dragonfly~44 exist in a stellar mass regime where core formation is expected to be maximally efficient \citep{DiCintio2014}. In contrast, the more massive galaxies of similar GC richness are more likely to have dark matter cusps which will increase their central dynamical masses in relation to a cored dark matter halo of similar mass.

Alternatively, it may be tempting to ascribe any differences between VCC~1448 and other galaxies to an over-estimation of VCC~1448's GC numbers by \citet{Lim2020} as it was measured with ground-based imaging. It may be possible that the ground-based GC number by Lim et al. is a severe overestimate. However, included in \citet{Lim2020} was the galaxy NGVSUDG-04/VLSB-B for which a total GC system of 13 $\pm$ 6 GCs was measured. \citet{Toloba2023} have now spectroscopically confirmed 14 GCs to be associated with this galaxy, giving some credence to the reliability of the \citet{Lim2020} methodology.

Further to this, we note that the lower limit for its GC content comes from the \textit{HST}/WFPC2 imaging of \citet{Seth2004}. Here the estimate of 34 GCs was determined from imaging corrected for luminosity incompleteness, but not for radial incompleteness \citep{Seth2004}. Specifically, the \citet{Seth2004} HST imaging only extends to $\sim1~R_{\rm e}$~and \textit{``excludes much of the halo light and any halo clusters that may exist"}. The addition of a radial correction to their total number can easily double the total number of GCs inferred for the galaxy since the GC half-number radius is frequently greater than the host galaxy's half-light radius (\citealp{Forbes2024}; Janssens et al. Submitted). At the very least, any moderate radial correction of GC numbers to the \citet{Seth2004} value will cause VCC~1448 to have high GC numbers for the dynamical mass we measure. This would then not affect the conclusions we draw from Figure \ref{fig:Ngc_mdyn}. It is also unable to reduce the predicted total halo mass for VCC~1448 given the \citet{Burkert2020} relationship sufficiently to alter the conclusions we draw from Figure \ref{fig:mass_profile}. 

Finally, it is worth considering VCC~1448 in comparison to Dragonfly 44 in Figure \ref{fig:Ngc_mdyn}. Many of our conclusions that hold for VCC~1448 also hold for Dragonfly 44. Namely, it also has higher GC counts for its dynamical mass than many previously studied galaxies \citep{Harris2013}. A partial explanation for this is Dragonfly 44 having a cored halo mass profile \citep{vanDokkum2019b} which would lead to a lower $V_{\rm rms}$ at fixed halo mass (GC number). Indeed we note that the large scatter of the plotted sample in Figure \ref{fig:Ngc_mdyn} is likely caused by a combination of the scatter in the GC number -- halo mass relationship and a scatter in the intrinsic velocity curves of these dark matter halos at fixed halo mass (GC number). Dragonfly 44 and VCC~1448 may thus be some of the most divergent dwarfs in what is referred to by many as a diversity of rotation curves problem (c.f., \citealp{Sales2022}) in the dwarf galaxy stellar mass regime. Similar conclusions have been reached for a larger sample of large half-light radius, GC-rich dwarfs in \citet{Forbes2024}.

\subsection{Kinematics \textit{vs.} Ellipticity}

An interesting difference between VCC~9, Dragonfly~44 and VCC~1448 is the level of rotation in their kinematics. We investigate this further in Figure \ref{fig:Vsigmavsellipticity}. Here we plot galaxies in the commonly used $\frac{\langle V \rangle}{\langle \sigma \rangle}$~(left axis) and $\lambda_{R}$~(right axis) \textit{vs.} ellipticity parameter spaces to further investigate their level of ordered vs. random motions (see e.g., \citealp{Cappellari2007}). 


In Figure \ref{fig:Vsigmavsellipticity} we include an isotropic line in $\frac{\langle V \rangle}{\langle \sigma \rangle}$ -- ellipticity space. This line is based on the tensor virial theorem and assumes a global anisotropy parameter (commonly $\delta$) of 0 (c.f. sec. 4.8.3 of \citealp{GalacticDynamics} or \citealp{Binney2005}). This is derived for edge-on systems. We show the intrinsic values of VCC~1448 that would cause us to produce the measurements in this work if viewed at a viewing angle of 75, 60, 45, and 36 degrees as the smaller connected stars. Any inclination effect where a galaxy is moved from a face-on-view to a more edge-on view will tend to move galaxies to the right in this parameter space \citep{GalacticDynamics}. This is also the direction of increasing global anisotropy. As such, any galaxy that is incorrectly assumed to be edge-on will systematically also reside in regions of lower anisotropy. There exists the possibility that our assumption of viewing VCC~1448 edge-on will cause us to underestimate the global anisotropy of the galaxy.

Additionally, in Figure \ref{fig:Vsigmavsellipticity} we include the commonly used parameter $\lambda_{\rm R}$ that quantifies the galaxy kinematic support. We convert our $\frac{\langle V \rangle}{\langle \sigma \rangle}$ to $\lambda_{\rm R}$ using the empirical relations derived in \citet{Emsellem2011} and listed by \citet{Penoyre2017}. We also include the divide between the regions prescribed for fast and slow central rotators as proposed by \citet{Emsellem2011} and updated in \citet{Cappellari2016}. Both VCC~9 and VCC~1448 are fast rotators centrally while Dragonfly 44 is a slow rotator. 

VCC~1448 lies extremely close to the isotropic line hence, there is some indication that the flattening of VCC~1448's morphology can be fully explained by its rotation alone. This contrasts with the commonly used explanation for (the majority) of larger elliptical galaxies that lie to the right of this line where the commonly assumed explanation is that there exists some level of anisotropy in the stellar orbits within the galaxy caused by its formation via (dry) mergers \citep{Naab2006}. However, this may become more complex when the exact details of the merger (see e.g., \citealp{Bois2011}) are considered. VCC~1448 may therefore be being viewed, at least partially, face-on. This would allow a level of anisotropy in its kinematics which may reflect a formation via galaxy mergers. As previously discussed, the conclusion that VCC~1448 may not be edge on does not affect the conclusions we draw for VCC~1448 in Figure \ref{fig:Ngc_mdyn}.

While of slightly lower stellar mass than the lowest mass bin in \citet{FraserMckelvie2022}, VCC~1448 has a log($V/\sigma$)$\approx-0.4$~which makes it extremely unlikely that they would classify it as a dynamically cold disk. The implication is that VCC~1448 cannot be easily explained as a rotationally supported, dynamically cold disk, nor as a dispersion-dominated slow rotator, but it is instead classified as an ``intermediate" system. As suggested by \citet{FraserMckelvie2022} this may imply that VCC~1448 either did not have the opportunity to accrete extra high-momentum gas from its surroundings or that it has an overly active merger history. While there is some evidence that the galaxy may not have had the opportunity to accrete high-momentum gas given its location in a cluster environment, here we prefer the merger explanation. VCC~1448 exhibits isophotal twists along with a lower dynamical mass than expected for estimations of its halo mass given either relationship, with a possible cause being a lack of equilibrium in the galaxy. These both are evidence that better aligns with the merger hypothesis than the high-angular momentum gas accretion hypothesis.

However, mergers and other tidal effects are not the only explanation for VCC~1448's kinematics. The simulations of \citet{CardonaBarrero2020} suggest a broad distribution in stellar kinematics for such large dwarf galaxies. Furthermore, the observational work of \citet{deLosReyes2023} has shown that dwarfs with kinematics similar to VCC~1448 may form in isolated void environments, where tidal effects and mergers play a much lesser role. Galaxies that form earlier and quench earlier likely have lower levels of $V/\sigma$ representative of their fast formation with an inability to readily acquire gas angular momentum. As discussed in \citet{FerreMateu2023} and \citet{Forbes2024}, Dragonfly~44 is likely a ``failed galaxy'', one that quenched early and has experienced little growth in stellar mass since then. We see some possible evidence of this whereby Dragonfly~44 has a much lower $V/\sigma$ than VCC~1448 which has a relatively younger age of $\sim7$~Gyr. 

It is worth noting that these formation scenarios do not need to be considered for VCC~9. Here \citet{FraserMckelvie2022} would classify it as a ``fast rotator" suggesting it has likely formed via the accretion of gas similar to the more commonly studied higher mass galaxies. Indeed studies such as \citet{Scott2020} suggest both of our Virgo cluster galaxies may exhibit a high level of rotation for a cluster dwarf (see their figs. 9/10). If both galaxies were relatively late to fall into the Virgo cluster (as will be seen later in Section \ref{sec:biasing}) some of their large size may be related to not yet experiencing the transformation process that many classical dE's have undergone in the cluster environment \citep{Toloba2015, Bidaran2020}. This would also imply that they were not pre-processed in a large group prior to infall.

Additionally, based on the kinematic evidence we suggest it is unlikely that an evolutionary link may exist between Dragonfly 44 and VCC~1448. In short, it is unclear how the passive evolution of VCC~1448 in the cluster environment would have enough impact on its kinematics to turn it into a slow rotator like Dragonfly 44. Conversely, our conclusion that VCC~1448 may have experienced past mergers makes it unlikely that it can represent a different evolutionary pathway for a `failed galaxy'' such as Dragonfly 44. 


Finally, the simulations of \citet{Pfeffer2024} suggest that dwarf galaxies in the mass regime of those studied here may be biased to be more GC-rich based on their initial gas properties. Higher-pressure gas lends itself to more efficient cluster formation (i.e., higher GC richness) and tends to create galaxies that are more pressure-supported. As such there may be an expectation that galaxies that are GC-rich should be preferentially pressure-supported and GC-poor galaxies being rotationally supported. VCC~9, VCC~1448 and Dragonfly~44 follow this trend, as their relative GC-richness (i.e., $M_{\rm GC}/M_\star$) increases (see Table \ref{tab:target_Summary}) they also have lower levels of $V/\sigma$. We note however that our sample is relatively small with only 3 galaxies of which 1 is at a slightly different stellar mass. A future study of a larger sample looking to probe these trends is vital to confirm their hypothesis.

\subsection{Globular Cluster System Biasing} \label{sec:biasing}

In Figure \ref{fig:rhee_diagram} we explore the idea of `biasing' which is that at fixed stellar mass, galaxies that fall in earlier have formed faster and thus have a higher fraction of their stellar mass bound in a richer GC system. Commonly this is observed as both the alpha enhancement and the cluster-centric distance/phase space position of galaxies correlating with their specific frequency (see e.g., \citealp{Liu2016, Mistani2016, Carleton2021}). Figure \ref{fig:rhee_diagram} displays a phase space diagram for the cluster environment. It is colour-coded based on the regions prescribed by the simulations of \citet{Rhee2017}. Darker shaded areas correspond to regions where galaxies predominantly fall in at earlier times than those in lighter regions. In addition to plotting the three target galaxies of our study, we include the Virgo cluster dwarf galaxy sample of \citet{Liu2016} which is comprised primarily of more compact, classical dE's in a similar stellar mass range to our targets. In the \citet{Liu2016} data, a clear trend was found for GC-richness and cluster-centric distance. 

As has been noted for other GC-rich UDGs, including Dragonfly~44, it is not clear UDGs follow this trend \citep{Gannon2023a, Forbes2023, FerreMateu2023, Toloba2023}. For example, the recessional velocity of Dragonfly~44 has led some authors to argue it may be part of a low mass galaxy group only currently being accreted to the Coma Cluster \citep{Alabi2018, vanDokkum2019b, Villaume2022}, which fits with its phase space positioning in Figure \ref{fig:rhee_diagram}. VCC~1448 resides in the late infall region of Figure \ref{fig:rhee_diagram} suggesting it is not an early infall into the galaxy cluster. We have further evidence for this from VCC~1448's stellar populations, which suggest a mass-weighted age of $\sim 7$ Gyr (Ferre-Mateu et al. in prep). A full stellar population analysis of VCC~1448's radial profile will be the subject of future work. VCC~1448 also resides in a region of phase space that is outside of the sample of \citet{Liu2016} that was used to argue for this `biasing' of dwarf galaxies in clusters. Similar conclusions have been reached by \citet{Toloba2023} who found that multiple large, GC-rich Virgo dwarfs are known to exist in this region of large velocities relative to the cluster (see their fig. 10). 

In Figure \ref{fig:rhee_diagram} VCC~9 resides in a region of mixed infall times making its fall in time ambiguous. We note that it does reside near the region traced by the \citet{Liu2016} Virgo dwarfs. In comparison to these Virgo dE's a relatively low GC specific frequency would be predicted which is in agreement with our observations. 

We conclude that while many dwarf galaxies may experience an elevated GC-richness (i.e., a higher than normal $M_{\rm GC}/M_\star$) due to their fast formation and early accretion into the cluster, this is not true for either Dragonfly~44 or VCC~1448. They likely have formed their high $M_{\rm GC}/M_\star$ external to the Coma and Virgo clusters respectively. As such, the special star formation environment common in a high redshift proto-cluster that leads to an elevated $M_{\rm GC}/M_\star$ may also be possible in field environments. Alternatively, a separate formation pathway to elevate the relative stellar mass in the GC system may be required for the galaxies forming in the field. 

\section{Conclusions} \label{sec:conclusions}
In this work, we have presented new Keck/KCWI spatially resolved data for the galaxy VCC~1448 along with new Keck/KCWI integrated data for the galaxy VCC~9. VCC~1448 is similar in size and magnitude to VCC~9 but hosts a far richer GC system. Our main conclusions based on these data are as follows:

\begin{itemize}
    \item We revise the recessional velocity of VCC~1448 to 2280$\pm$6 \kms. We suggest the previous value, which is based on an HI detection, may be inaccurate due to ram pressure stripping of gas in the cluster environment.
    \item We present a radial kinematic profile for VCC~1448 to just beyond the half-light radius. VCC~1448 has an approximately flat velocity dispersion profile, however, it exhibits a rising rotational component. 
    \item We confirm 10 compact sources within our field of view are GCs associated with VCC~1448, 1 GC associated with VCC~9 and 1 that is likely a foreground star. For VCC~1448, we note that the portion of the GC system that we have sampled has noticeably higher recessional velocity than the galaxy's main stellar body. This is at least partially (if not fully) due to the bias in spatial sampling of our data to an area of the galaxy with rising velocity. 
    \item We derive a dynamical mass from VCC~9's stellar velocity dispersion and rotation of 2.69$\pm$2.29$\times$10$^9$~M$_{\odot}$ within 3.96 kpc. It is in agreement with halo profiles of total mass as expected from the GC number -- halo mass relationship along with the stellar mass -- halo mass relationship. 
    \item Based on our velocity measurements we can produce two independent mass estimates for VCC~1448 at different radii, building a crude mass profile. One mass estimate comes from the stellar velocity dispersion and its rotation of 1.96$\pm$0.68$\times$10$^9$~M$_{\odot}$ within 4.67 kpc. The other mass estimate comes from the velocities of the GC system of 8.55$^{+1.73} _{-0.77}$$\times$10$^9$~M$_\odot$ within 6.48 kpc. Our mass profile is in better agreement with the halo profiles of total mass based on the stellar mass -- halo mass relationship than the GC number -- halo mass relationship.
\end{itemize}

Further to this, we made comparisons to the galaxy Dragonfly~44 which is similar in both size and GC-richness to VCC~1448 but has a lower stellar mass than either VCC~9 or VCC~1448. Based on a comparison of the three galaxies we conclude:

\begin{itemize}
    \item When considering the position of the three galaxies $N_{\rm GC}$ -- $M_{\rm Dyn}$ space, VCC~1448 and Dragonfly~44 exhibit rich GC numbers for their dynamical mass. VCC~9 however follows the normal galaxy sample. If each galaxy's total halo mass estimate from their GC counts is correct, this suggests a large diversity in dwarf galaxy rotation curves in this stellar mass regime (i.e., $\log(M_{\star} / M_\odot)\approx 8-9$).  
    \item The position of VCC~1448 in $\frac{\langle V \rangle}{\langle \sigma \rangle}$ -- ellipticity space may suggest that we are not viewing the galaxy edge-on. While this does not change our previous conclusions it suggests a level of anisotropy that may support the hypothesis of a formation via mergers. A formation via mergers is further supported by the isophotal twists found in other studies.
    \item VCC~9 exhibits more rotational support than VCC~1448 which in turn exhibits more rotational support than Dragonfly 44. This combined with their different GC systems suggests multiple formation pathways are needed to create a large dwarf galaxy at fixed luminosity. It also supports simulated predictions where GC formation efficiency is affected by the gas properties during initial star formation.
    \item Dragonfly 44's position in stellar mass -- halo mass space appears to require early quenching and passive evolution. It is not clear how the passive evolution of VCC~1448 could alter its kinematics into those seen in Dragonfly 44. This evolutionary link between Dragonfly 44 and VCC~1448 appears unlikely. 
\end{itemize}

\section*{Acknowledgements}
We thank the anonymous referee for their careful, constructive consideration of our manuscript. We thank E. Toloba and T. Lisker for providing data to aid comparisons to her work. JSG thanks C. Johnson for sharing their wisdom throughout the creation of this work, she will be deeply missed. Some of this work was completed while JSG was an ECR visiting fellow at the Institutio de Astrofisica de Canarias, he is grateful for their support. This research was supported by the Australian Research Council Centre of Excellence for All Sky Astrophysics in 3 Dimensions (ASTRO 3D), through project number CE170100013. DF, WC and JB thank ARC DP220101863 for financial support. AJR was supported by the National Science Foundation grant AST-2308390. This work was supported by a NASA Keck PI Data Award, administered by the NASA Exoplanet Science Institute. AFM has received support from RYC2021-031099-I and PID2021-123313NA-I00 of MICIN/AEI/10.13039/501100011033/FEDER,UE, NextGenerationEU/PRT. The data presented herein were obtained at the W. M. Keck Observatory, which is operated as a scientific partnership among the California Institute of Technology, the University of California and the National Aeronautics and Space Administration. The Observatory was made possible by the generous financial support of the W. M. Keck Foundation. The authors wish to recognize and acknowledge the very significant cultural role and reverence that the summit of Maunakea has always had within the indigenous Hawaiian community. We are most fortunate to have the opportunity to conduct observations from this mountain.

\section*{Data Availability}
The KCWI data presented are available via the Keck Observatory Archive (KOA) 18 months after observations are taken.



\bibliographystyle{mnras}
\bibliography{bibliography} 




\appendix

\section{Example GC Spectra}
\begin{figure*}
    \centering
    \includegraphics[width = 0.95\textwidth]{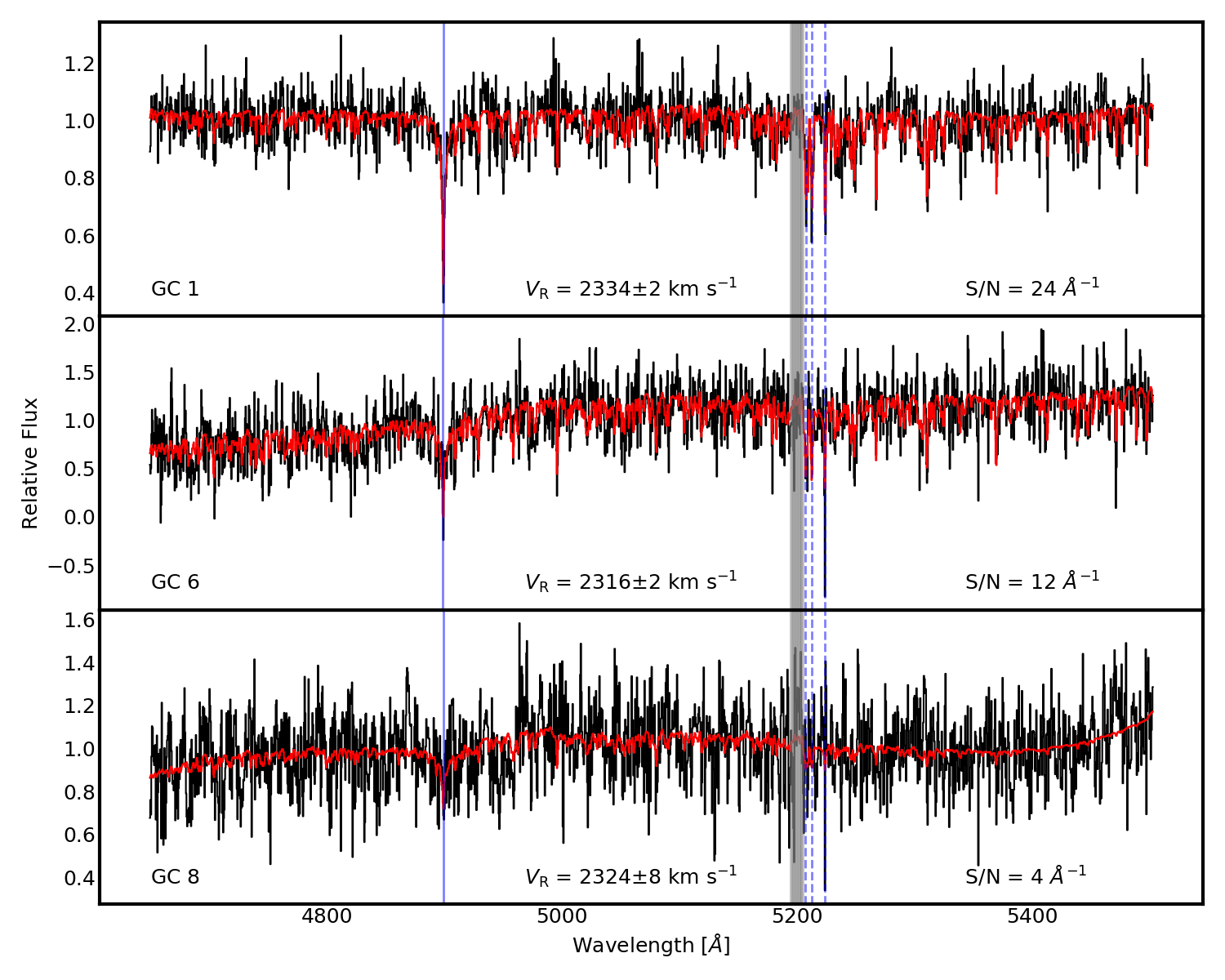}
    \caption{Three example GC spectra that are analysed in this work. From top to bottom, these spectra are for GC 1 (S/N = 24 \AA$^{-1}$), GC 6 (S/N = 12 \AA$^{-1}$) and GC 8 (S/N = 4 \AA$^{-1}$) as indicated on the plot. In black we plot the spectra with best fitting \texttt{pPXF} fit in red. Blue vertical lines indicate the position of H$\beta$~(solid line) and the Mg$b$~triplet (dashed lines) at the measured redshift. The grey fill shows the expected location of any residuals from the prominent 5199\AA~sky feature. We conclude our fits to these data are reasonable. These fits are typical and other fits of our data are of similar quality.}
    \label{fig:gc_spectra}
\end{figure*}

This appendix provides Figure \ref{fig:gc_spectra} with 3 example spectra from our GC fitting.


\bsp	
\label{lastpage}
\end{document}